\documentclass[fleqn]{2023SCGE}
\usepackage{float}
\usepackage{natbib}
\usepackage{graphicx}
\usepackage{appendix}

\newcommand\ion[2]{\text{#1\,\textsc{\lowercase{#2}}}}	
\newcommand{\HI}{\ion{H}{I}}
\newcommand{\HIFAST}{\texttt{HiFAST}}

\begin{document}

\ensubject{subject}

\ArticleType{Article}
\SpecialTopic{SPECIAL TOPIC: }
\Year{2023}
\Month{?}
\Vol{?}
\No{?}
\DOI{??}
\ArtNo{000000}
\ReceiveDate{?}
\AcceptDate{?}

\title{\HIFAST: an \HI\ data calibration and imaging pipeline for FAST}

\author[1]{Yingjie Jing\thanks{jyj@nao.cas.cn}\thanks{FAST fellow}}{}
\author[1,2,3]{Jie Wang}{}
\author[1,3]{Chen Xu}{}
\author[1,3]{Ziming Liu}{}
\author[1,3]{Qingze Chen}{}
\author[1,3]{Tiantian Liang}{}
\author[1,6]{Jinlong Xu}{}
\author[4]{\\Yixian Cao}{}
\author[5]{Jing Wang}{}
\author[3,1]{Huijie Hu}{}
\author[1,6]{Chuan-Peng Zhang}{}
\author[1,2,3]{Qi Guo}{}
\author[1,2,3,7]{Liang Gao}{}
\author[1,6,8]{Mei Ai}{}
\author[1,8]{\\Hengqian Gan}{}
\author[1,8]{Xuyang Gao}{}
\author[1,8]{Jinlin Han}{}
\author[1]{Ligang Hou}{}
\author[1,3]{Zhipeng Hou}{}
\author[1,6]{Peng Jiang}{}
\author[9,10]{Xu Kong}{}
\author[9,10]{\\Fujia Li}{}
\author[1,3]{Zerui Liu}{}
\author[1]{Li Shao}{}
\author[11]{Hengxing Pan}{}
\author[1,12]{Jun Pan}{}
\author[1,8]{Lei Qian}{}
\author[1]{Jinghai Sun}{}
\author[13]{\\Ningyu Tang}{}
\author[1]{Qingliang Yang}{}
\author[1,8]{Bo Zhang}{}
\author[14,15]{Zhiyu Zhang}{}
\author[1,6,8]{Ming Zhu}{}

\AuthorMark{Jing}

\AuthorCitation{....}

\address[1]{National Astronomical Observatories, Chinese Academy of Sciences, Beijing 100012, China}
\address[2]{Institute for Frontiers in Astronomy and Astrophysics, Beĳing Normal University, Beĳing 102206, China}
\address[3]{School of Astronomy and Space Science, University of Chinese Academy of Sciences, Beijing 10049, China}
\address[4]{Max Planck Institute for Extraterrestrial Physics, Garching 85748, Germany}
\address[5]{Kavli Institute for Astronomy and Astrophysics, Peking University, Beijing 100871, China}
\address[6]{Guizhou Radio Astronomical Observatory, Guizhou University, Guiyang 550000, China}
\address[7]{Institute of Computational Cosmology, Department of Physics, University of Durham, Durham DH1 3LE, UK}
\address[8]{CAS Key Laboratory of FAST, National Astronomical Observatories, Chinese Academy of Sciences, Beijing 100101,
China}
\address[9]{ Deep Space Exploration Laboratory / Department of Astronomy, University of Science and Technology of China, Hefei 230026, China}
\address[10]{ School of Astronomy and Space Science, University of Science and Technology of China, Hefei 230026, China}
\address[11]{ Astrophysics, University of Oxford, Oxford OX1 3RH, UK}
\address[12]{ College of Earth Sciences, Guilin University of Technology, Guilin 541004, China}
\address[13]{ Department of Physics, Anhui Normal University, Wuhu, Anhui 241002, China}
\address[14]{ School of Astronomy and Space Science, Nanjing University, Nanjing 210023, China}
\address[15]{ Key Laboratory of Modern Astronomy and Astrophysics, Nanjing University, Nanjing 210023, China}


\abstract{The Five-hundred-meter Aperture Spherical radio Telescope (FAST) has the largest aperture and a 19-beam L-band receiver, making it powerful for investigating the neutral hydrogen atomic gas (\HI) in the universe. We present \HIFAST\ (\url{https://hifast.readthedocs.io}), a dedicated, modular, and self-contained calibration and imaging pipeline for processing the \HI\ data of FAST. The pipeline consists of frequency-dependent noise diode calibration, baseline fitting, standing wave removal using an FFT-based method, flux density calibration, stray radiation correction, and gridding to produce data cubes. These modules can be combined as needed to process the data from most FAST observation modes: tracking, drift scanning, On-The-Fly mapping, and most of their variants. With \HIFAST, the RMS noises of the calibrated spectra from all 19 beams were only slightly ($\sim 5$\%) higher than the theoretical expectation. The results for the extended source M33 and the point sources are consistent with the results from Arecibo. The moment maps (0,1 and 2) of M33 agree well with the results from the Arecibo Galaxy Environment Survey (AGES) with a fractional difference of less than 10 per cent. For a common sample of 221 sources with signal-to-noise ratio S/N $>10$ from the Arecibo Legacy Fast ALFA (ALFALFA) survey, the mean value of fractional difference in the integrated flux density, $S_{\mathrm{int}}$, between the two datasets is approximately 0.005 per cent, with a dispersion of 15.4 per cent. Further checks on the integrated flux density of 23 sources with seven observations indicate that the variance in the flux density of the source with luminous objects ($S_\mathrm{int}$ $ > 2.5$ Jy km s$^{-1}$) is less than 5 per cent. Our tests suggest that the FAST telescope, with the efficient, precise, and user-friendly pipeline \HIFAST, will yield numerous significant scientific findings in the investigation of the \HI\ in the universe.}



\maketitle


\begin{multicols}{2}

\section{Introduction}
The Five-hundred-meter Aperture Spherical radio Telescope (FAST) is the largest single-dish and most sensitive radio telescope in the world \citep[][]{Nan2011}. It has an illuminated aperture of 300\,m and a 19-beam L-band receiver that covers frequencies from 1050 to 1450 MHz \citep[e.g.][]{Jiang2019,Jiang2020,Qian2020,Yin2023}, making it a powerful instrument for observing the neutral hydrogen atomic gas (\HI\ lines). The raw spectral data must be calibrated in terms of amplitude and transformed by frequency before any scientific analysis or comparison with other telescopes can be conducted. In this paper, we introduce \HIFAST\footnote{\url{https://hifast.readthedocs.io}}, a modular and flexible \HI\ data calibration and imaging pipeline for FAST.

The paper is organised as follows. We describe the pipeline workflow in Section 2. The calibration procedures, including temperature and flux calibration, are presented in Section 3. The Radio Frequency Interference (RFI) flagging and Doppler correction are described in Sections 4 and 5. The imaging procedures are described in Section 6. A pipeline verification test on FAST data is presented in Section 7. The code implementations are described in Section 8, and a concluding summary is provided in Section 9. More details and tests on the module flux calibration, standing wave removal, and stray radiation correction are extensively described in separate papers: Liu et al., Xu et al., and Chen et al. (in preparation, hereafter referred to as papers II, III, and IV, respectively).

\section{Scheme of \HIFAST}

\begin{figure*}
        \centering
	\includegraphics[width=1.8\columnwidth]{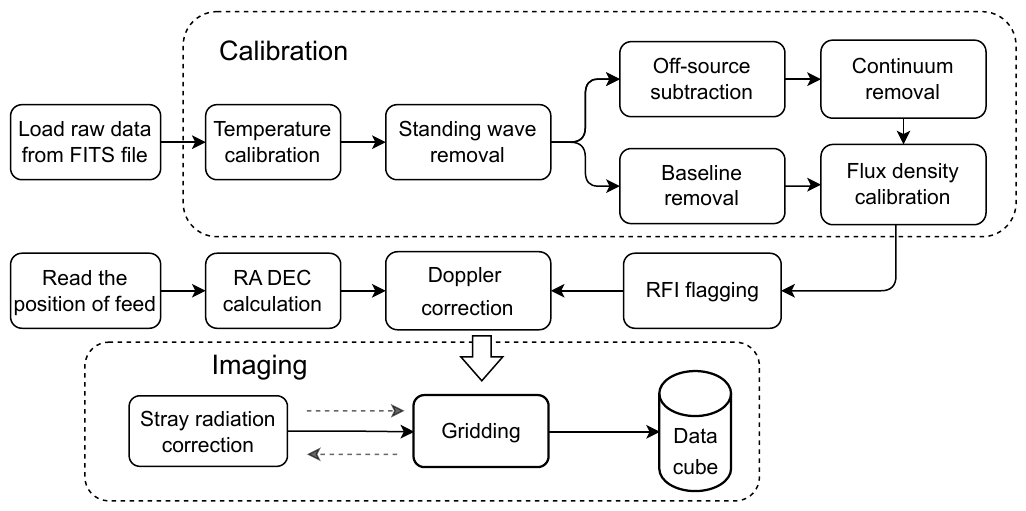}
	\caption{Flowchart of the typical \HIFAST\ calibration and imaging process.}
	\label{fig:flow}
\end{figure*}

\HIFAST\ supports the processing of the data observed in various modes, including tracking, drifting scan, and On-The-Fly (OTF) mapping, as well as most of their variants \footnote{\url{https://fast.bao.ac.cn/cms/article/24/}}  \citep[see more details on FAST observation modes in][]{Jiang2020}.

In Fig.~\ref{fig:flow}, we present a typical calibration and imaging workflow in \HIFAST. The raw spectral data recorded by FAST are stored in the Flexible Image Transport System (FITS) files, mainly including sky frequency and uncalibrated intensity (power) in each spectral channel recorded as single-precision floating point data. Additionally, the trajectory of the phase centre of the feed is stored in a corresponding Office Open XML Workbook (.xlsx) file. \HIFAST\ loads the raw spectral data from the FITS files and calibrates them to the temperature using a noise diode as reference power input. To obtain the antenna temperature of the \HI\ line, the standing wave is first removed, followed by the off-source subtraction or the removal of the baseline to eliminate the system temperature. The temperature is then calibrated to the flux density. Meanwhile, \HIFAST calculates the right ascension (RA) and the declination (DEC) of each spectrum using the three-dimensional coordinates of the feed and the measurement time. RFI flagging and Doppler correction procedures are also included. For mapping observation modes, such as drifting scans or OTF, the imaging process consists of an iterative stray radiation correction and gridding. The final cube data is saved in a standard FITS file. These modules can be combined in different ways depending on user needs. Detailed descriptions of these modules are presented in the following sections.

\section{Calibration procedures}

\begin{figure*}
        \centering
	\includegraphics[width=2\columnwidth]{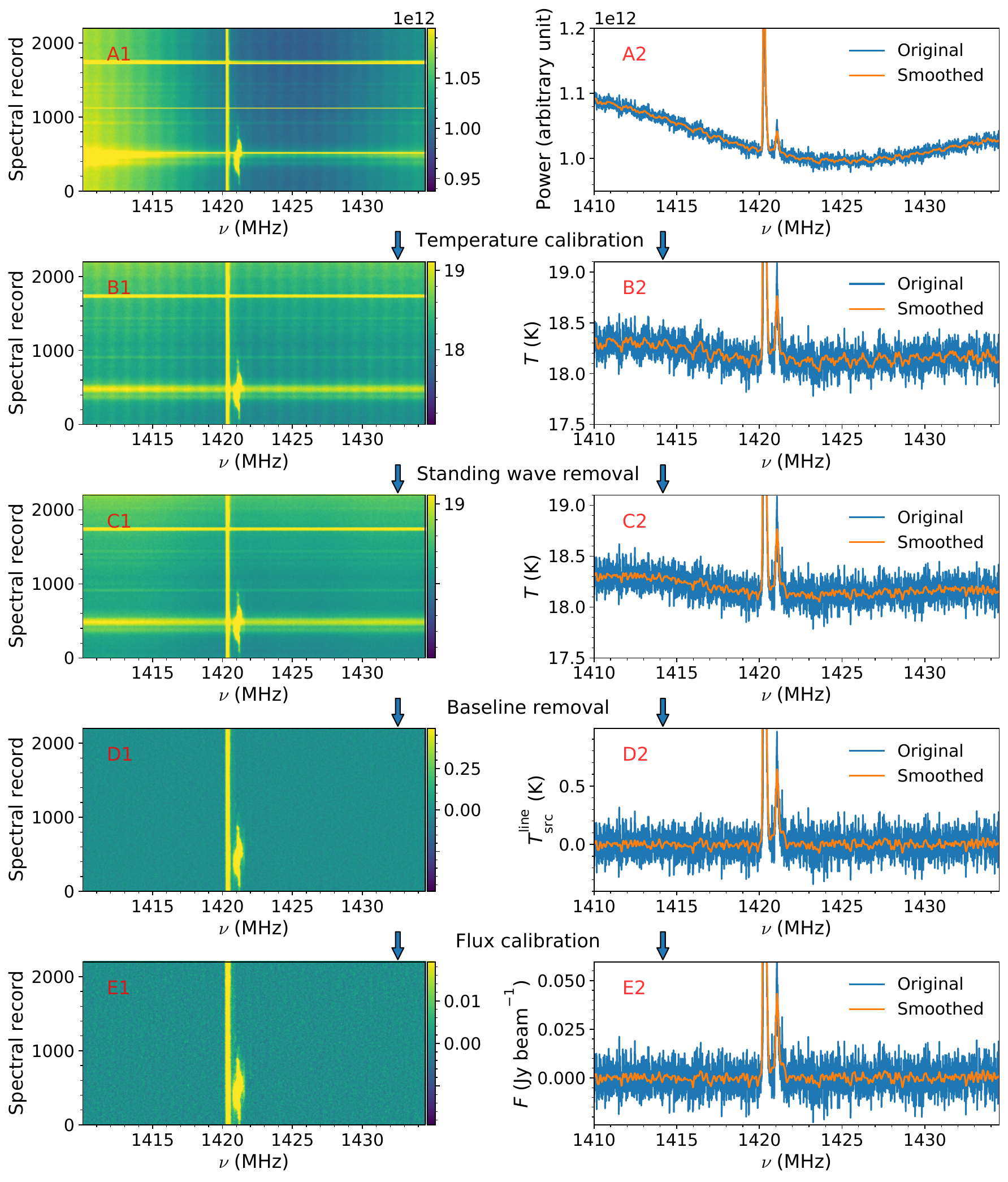}
	\caption{An example of typical calibration process for one polarization of spectra from one beam taken in a mapping scan. Left panels: waterfall plots, i.e., power (A), temperature (B, C and D) and flux density (E) as a function of frequency $\nu$ and spectral record number, for 2200 continuous spectral records. Each spectrum is recorded with a 0.5 second exposure time. Right panels: the average of 10 consecutive records (corresponding to 5 seconds of integration time, depicted in blue lines) selected from the left panels. The smoothed spectra are shown in orange lines for enhanced clarity. The arrows between the panels indicate the sequence of procedures: A$\rightarrow$B (refer to Section~\ref{sec:tcal}) involves calibrating the raw power data to the temperature of the spectra using the noise diode; B$\rightarrow$C entails the removal of standing waves via the FFT-filter method; C$\rightarrow$D involves baseline subtraction; and D$\rightarrow$E includes calibrating the temperature to the flux density using a stable standard flux calibrator.}
	\label{fig:waterfall}
\end{figure*}

The raw observation data records the detected power $P$ of the receiver output, comprising various contributions \citep[e.g.][]{Neil2002,Winkel2012,Jiang2020}. This can be expressed as:
\begin{align}
P &= gT\\
 &= g({T}_\mathrm{src} + {T}_\mathrm{sys}) \\
 &= g({T}_\mathrm{src} + {T}_{\mathrm{rec}}+ {T}_{\mathrm{sw}} + {T}_{\mathrm{gr}}+{T}_{\mathrm{atm}}+{T}_{\mathrm{bg}} \ [+ T_\mathrm{cal}]),
\end{align}
where $g$ represents the gain of the receiving system. $T_{\mathrm{src}}$ is the antenna temperature induced by the flux density of the astronomical source, including line profiles $T_{\mathrm{src}}^\mathrm{line}$ (emission or absorption) and a continuum contribution $T_{\mathrm{src}}^\mathrm{cont}$. $T_{\mathrm{sys}}$ is the system temperature that accounts for all other contributions. Specifically, $T_{\mathrm{rec}
}$ represents the receiver's noise temperature, $T_{\mathrm{sw}}$ is the contribution from standing waves, $T_{\mathrm{gr}}$ arises from ground radiation, $T_{\mathrm{atm}}$ is the atmospheric contribution, $T_{\mathrm{bg}}$ includes the cosmic microwave background and the Galactic emissions, and $T_{\mathrm{cal}}$ is the calibration noise injected by a noise diode for calibration purposes. The term $T_{\mathrm{cal}}$ only contributes when the noise diode is activated. Previous studies detail the signal path in the telescope \citep[e.g.][]{Jiang2019, Dunning2017, Jeganathan2017}.

For scientific analysis, the raw data are calibrated to transform power $P$ to temperature $T$ using a noise diode as reference power input (see Section \ref{sec:tcal}). Subsequently, the line temperature $T_{\mathrm{src}}^\mathrm{line}$ is derived (refer to Section \ref{sec:src}). A standard source with known flux density is used to convert the temperature to the flux density (see Section \ref{sec:flux}). This calibration is applied to each spectrum, obtained under different beams and polarizations (XX and YY). Fig.~\ref{fig:waterfall} illustrates this calibration for 2200 spectra in polarization mode XX. The following subsections describe each calibration stage in detail.

\subsection{Temperature calibration} \label{sec:tcal}
FAST utilises a noise diode with a known temperature (approximately 1 K or 10 K) to measure the gain ($g$) and calibrate the recorded power to temperature. The stability of the gain $g$ can be maintained for several minutes, with fluctuations on the order of a few percent for FAST \citep{Jiang2020}. Consequently, the noise diode needs to be periodically injected, such as for 1 second every 5 minutes. Within each injection period, the measured gain, denoted as $g_\mathrm{m}$, can be determined using the equation:

\begin{equation}
    g_\mathrm{m}(\nu) = \frac{[P_\mathrm{cal\_on} - P_\mathrm{cal\_off}^\mathrm{near}]^\mathrm{smo}(\nu)}{{T_\mathrm{cal}}^\mathrm{smo}(\nu)},
\end{equation}

Here, $T_\mathrm{cal}$ represents the pre-measured temperature of the noise diode\footnote{\url{https://fast.bao.ac.cn/cms/category/telescope_performance_en/noise_diode_calibration_report_en/}}, $P_\mathrm{cal\_on}$ corresponds to the power measured with the noise diode activated, and $P_\mathrm{cal\_off}^\mathrm{near}$ represents the power of the closest spectrum obtained with the noise diode deactivated, close in time to $P_\mathrm{cal\_on}$. The term ``smo" denotes the process of smoothing the frequency-dependent power and temperature of the noise diode along the frequency $\nu$ (i.e., channel) using a Gaussian filter or moving average. This smoothing reduces the variance induced by random noise. The typical smoothing width is 2 MHz, which covers 263 channels in wide-band data\footnote{\url{https://fast.bao.ac.cn/cms/article/26/}} with a frequency resolution of 7.6 kHz.

The calibration of power to temperature for the spectra obtained when the noise diode is deactivated is performed using the equation:

\begin{align}
T_\mathrm{cal\_off}(\nu) &= \frac{P_\mathrm{cal\_off}}{g_\mathrm{m}(\nu)}
\end{align}

For the spectra obtained when the noise diode is activated (with the contribution from the noise diode subtracted), the calibration of power to temperature is given by:

\begin{align}
T_\mathrm{cal\_on}(\nu) &= \frac{P_\mathrm{cal\_on}}{g_\mathrm{m}(\nu)} - {T_\mathrm{cal}}^\mathrm{smo}(\nu),
\end{align}

where $g_\mathrm{m}$ represents the nearest measured gain in time.

An example of the temperature calibration process is shown in the top two panels of Fig.~\ref{fig:waterfall}. The power of the spectrum is shown in the first rows, A1 and A2, while the calibrated temperature of the spectra is presented in the subsequent rows, B1 and B2.

\subsection{Source temperature} \label{sec:src}
The temperature $T$ obtained by gain ($g$) calibration is the composition of the target astronomical source temperature ${T}_{\mathrm{src}}$ and the system temperature ${T}_\mathrm{sys}$. ${T}_{\mathrm{src}}$ potentially includes line profiles $T_{\mathrm{src}}^\mathrm{line}$(emission or absorption) and a continuum contribution $T_{\mathrm{src}}^\mathrm{cont}$. In studies of \HI\, the primary objective is to determine the line profile contribution $T_{\mathrm{src}}^{\mathrm{line}}$.

To obtain $T_{\mathrm{src}}^\mathrm{line}$, there are two approaches that can be used. The first method involves measuring $T_\mathrm{sys}$ by observing a blank sky region as an off-source measurement \citep[e.g.][]{Neil2002}. This measurement is subtracted from the total temperature $T$. Subsequently, a polynomial baseline fitting can be applied to eliminate the continuum contribution (including $T_{\mathrm{src}}^\mathrm{cont}$ and potential changes in $T_\mathrm{sys}$) to extract line profiles. The second method involves fitting a baseline directly onto source-on spectra across frequencies and then subtracting it, effectively removing both $T_\mathrm{sys}$ and the continuum contribution.

However, both techniques have limitations. It can be difficult or even impossible to select off-source observations in certain cases, such as extended sources like M31 or the ubiquitous \HI\ lines of the Milky Way. Additionally, fitting the baseline can be a challenge when ${T}_\mathrm{sys}$ has a complex frequency dependence. To address these issues, \HIFAST\ has modules designed to handle these two methods, depending on the specific situation.

In particular, ${T}_{\mathrm{sw}}$, which is involved in these methods, varies slightly over time (see details in Fig.~\ref{fig:onoff} of Section \ref{sec:ref} and Paper III), leading to different ${T}_\mathrm{sys}$ values in the off-source and on-source observations. The primary standing wave in FAST exhibits a period of about 1 MHz, which is comparable to or smaller than the width of numerous source signals. Consequently, it cannot be eliminated during baseline fitting either. Therefore, a separate method is needed to eliminate it (Section \ref{sec:sw}).

In the following subsections, we describe the required modules separately in detail.

\subsubsection{Off-source subtraction} \label{sec:ref}

The estimation of the off-source spectrum is different in the tracking and mapping modes. We will explain these two approaches below.

\begin{figure}[H]
	\centering
        \includegraphics[width=1.\columnwidth]{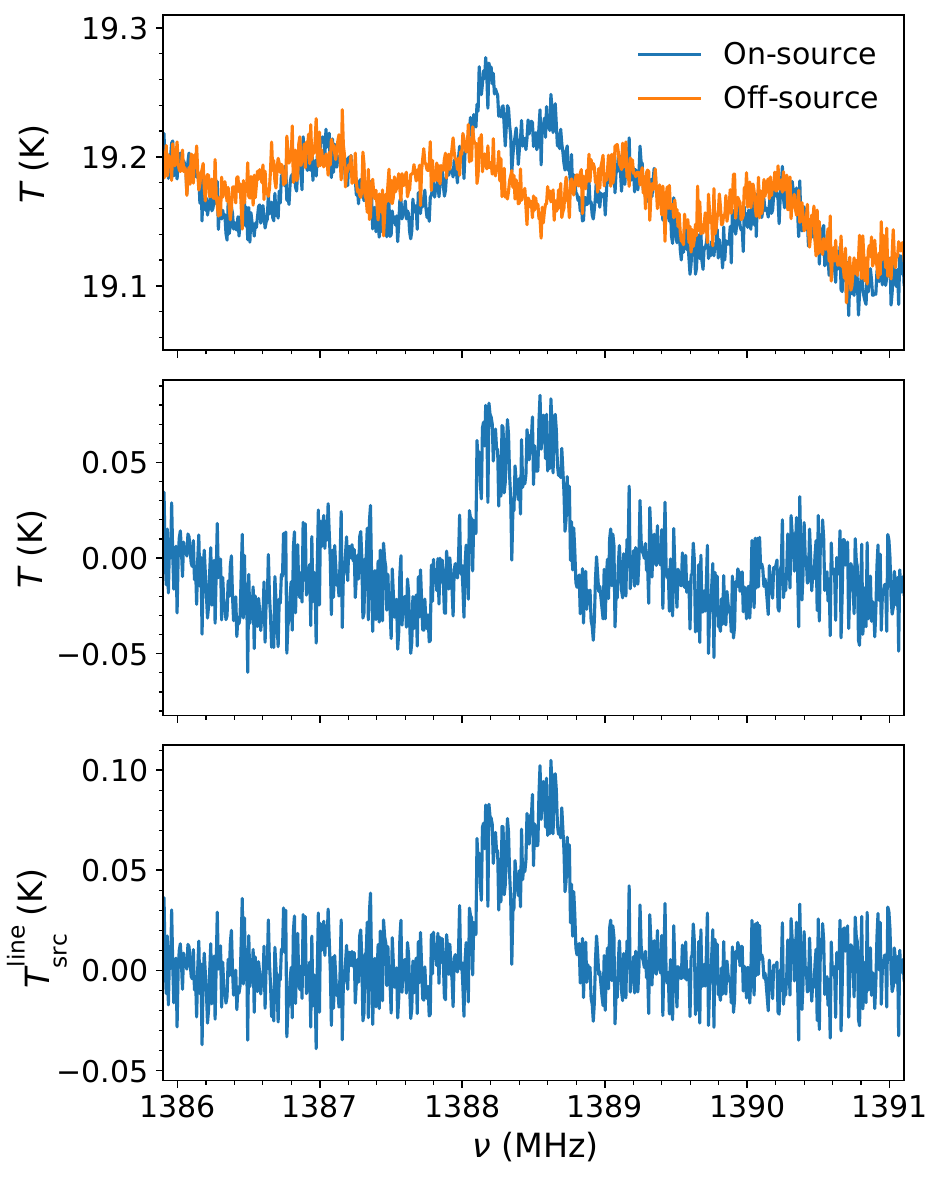}
	\caption{An example of off-source subtraction in position-switching tracking mode. The top panel shows the calibrated temperature of on-source (blue) and off-source (orange) observation, which is a blank sky region close to the source. The middle panel shows the target spectrum obtained by subtracting the off-source from the on-source spectrum. The bottom panel illustrates the result after removing the residual standing wave.}
	\label{fig:onoff}
\end{figure}

In tracking mode, position switching is commonly used to acquire blank sky information. This involves the telescope observing the target object for a certain amount of time and then moving to a blank sky region for an equal duration as an off-source observation. An example of this data processing is illustrated in Fig.~\ref{fig:onoff}. The top panel displays the temperature-calibrated spectrum of the on-source (blue) and off-source (orange) observations, with a ${T}_\mathrm{sys}$ of approximately $19.06$ K, including a sine standing wave. Additionally, an \HI\ emission line is present at a frequency of approximately $1388.5$ MHz. The middle panel shows the results of subtracting the off-source spectrum from the on-source spectrum. The temperature of the baseline is now close to zero, as most of the ${T}_\mathrm{sys}$ has been removed. However, a minor residual sine wave remains, caused by the phase shift of the standing wave with time. This can be eliminated using the sine fitting method described in Section~\ref{sec:sw}, as seen in the bottom panel. An alternative approach would be to remove the standing wave before performing the off-source subtraction, as the standing wave in the off-source observations differs from that in the on-source observations.

\begin{figure*}
	\centering
        \includegraphics[width=2.\columnwidth]{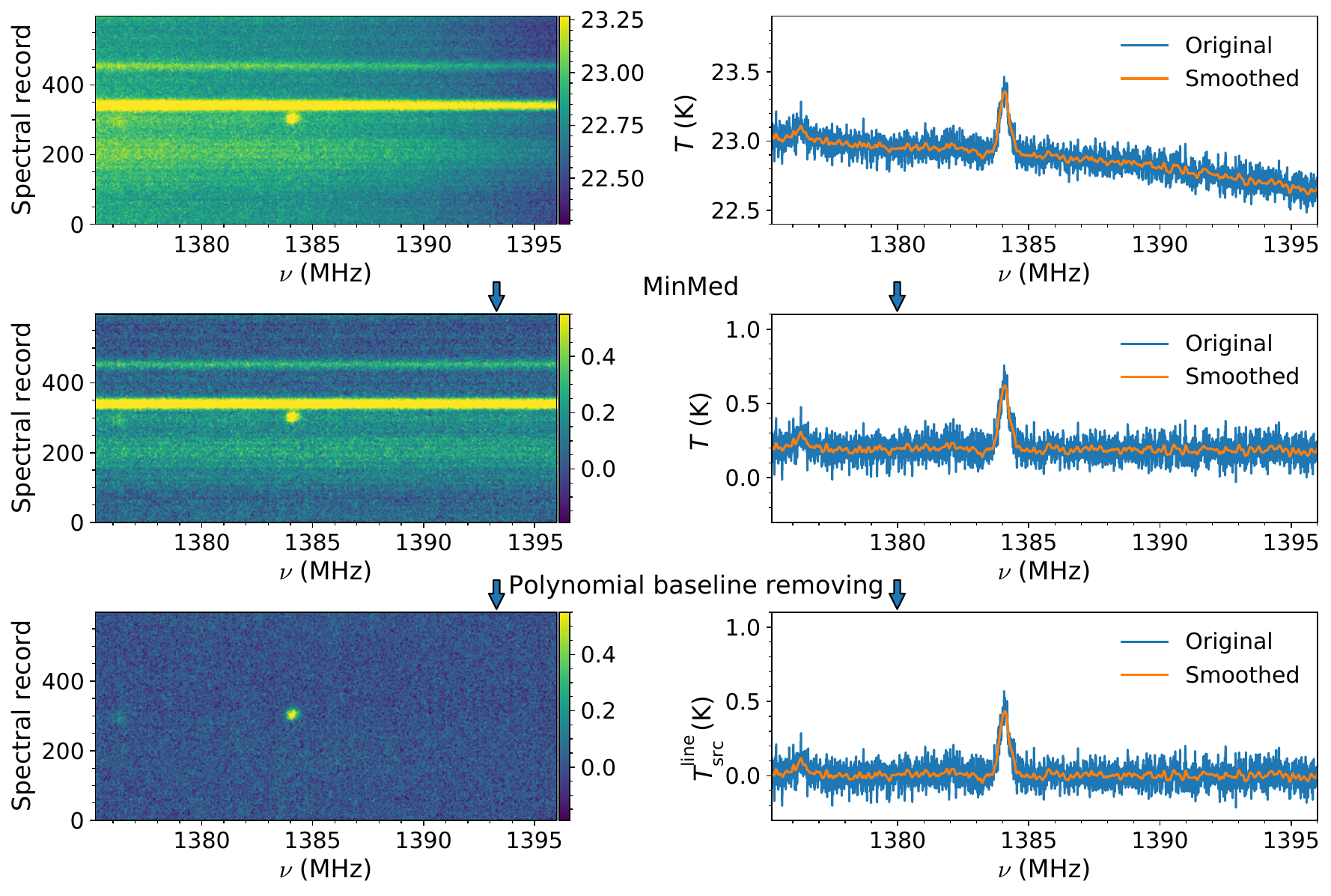}
	\caption{Similar to Fig.~\ref{fig:waterfall}, but for the ``MinMed" method used in off-source subtraction. Top panels: spectra with the standing wave removed; middle panels: spectra applied with the MinMed method; bottom panels: spectra applied with a lower-order polynomial baseline fitting to remove continuum contribution in the source and other possible origins.}
	\label{fig:MinMed}
\end{figure*}

In drifting scans and OTF mapping observations for point or small extended sources, off-source observation should be carefully selected \citep[e.g.][]{Neil2002, Wolfe2015} in general. The ``MinMed" method is usually used in this process. This method was introduced by \citet{Putman2002,Putman2003} and has been used in a number of investigations \citep[e.g.][]{Davies2011,Taylor2014}. The process involves dividing the scanned spectra into segments over time for each channel, calculating the median value for each segment, and then selecting the minimum of these medians as the off-source spectra. The median value is unaffected by the on-source spectra data when the angular size of the sources is sufficiently small compared to the area covered by each segment. Fig.~\ref{fig:MinMed} shows an example of data processing in this method, where the waterfall plot and the average of 10 consecutive records near the target source are shown in the left and right panels, respectively. Initially, we applied the ``MinMed'' method to estimate and subtract the off-source from temperature-calibrated spectra. The resulting waterfall plot for a portion of the scanning data is displayed in the left panel. Notably, a source was present at a frequency of approximately 1384 MHz, located near the 325th spectral record. In addition, a few spectra with elevated continuum levels are present around the 329th and 340th spectral records. To further refine the data, we removed a lower-order polynomial baseline from each spectrum using the method described in Section \ref{sec:bld}. The refined result can be found in the bottom panels of Fig.~\ref{fig:MinMed}.

\subsubsection{Baseline fitting} \label{sec:bld}

\begin{figure*}
        \centering
	\includegraphics[width=2\columnwidth]{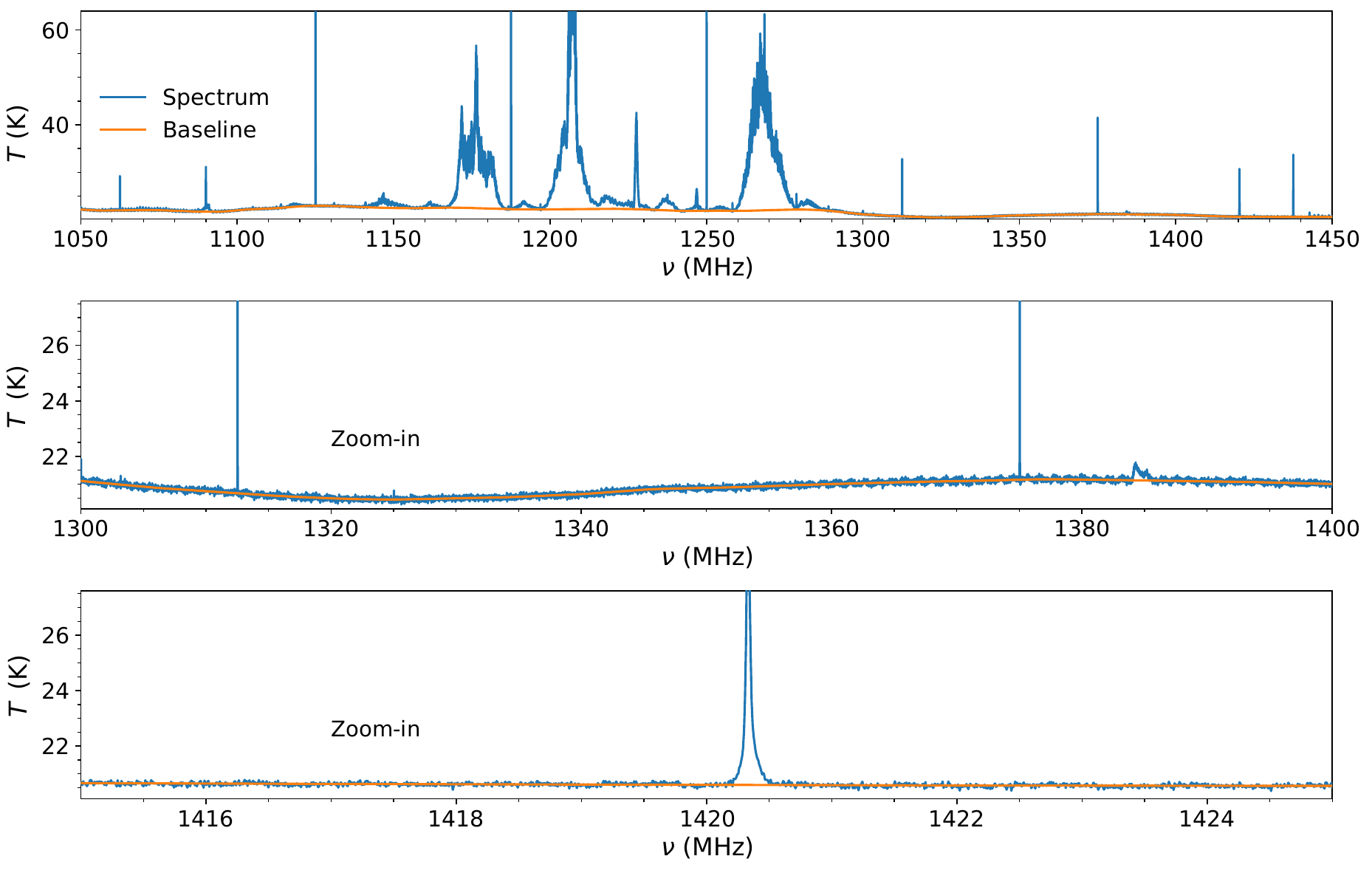}
	\caption{Baseline fitting result using arPLS. The blue and orange lines represent the spectrum and the fitted baseline, respectively. The upper panel displays the complete spectrum, while the middle and bottom panels provide a zoomed-in view.}
	\label{fig:bl}
\end{figure*}

The baseline fitting process involves isolating the line profile $T_{\mathrm{src}}^\mathrm{line}$ by treating the continuum contribution $T_{\mathrm{src}}^\mathrm{cont}$ and the system temperature component $T_\mathrm{sys}$ as the baseline. To obtain an accurate line profile, the baseline must be fitted in a signal-free region and then removed from the target spectrum. In drifting scan and OTF modes, manually selecting a signal-free region for baseline fitting can be difficult and time-consuming, especially when dealing with numerous spectra. To address this issue, the arPLS \citep[asymmetrically reweighted penalized least squares smoothing;][]{Baek2015} method is employed for automatic baseline fitting without having to select a signal-free region beforehand. The arPLS method is a type of penalized least squares method \citep[e.g.][]{Eilers2003, Zhang2010} and is effective even in the presence of small noise in the data.

In the arPLS method, the baseline $z$ of a spectrum with data points $\mathbf{y}$ is obtained using the following equation:

\begin{equation}
\mathbf{z}=\left
(\mathbf{W}+\lambda \mathbf{D}^T\mathbf{D}\right)^{-1} \mathbf{W} \mathbf{y}, \label{eq:bld-PLS}
\end{equation}
where $\mathbf{D}$ is the second-order difference matrix, $\lambda$ is a parameter that adjusts the balance between $\mathbf{y}$ and $\mathbf{z}$, and $\mathbf{W}$ is a diagonal matrix with weight vector $\mathbf{w}$ of $\mathbf{y}$ on its diagonal. The arPLS method iteratively adjusts and applies weights to the data points, refining the baseline until the weight changes become negligible or fall below a predetermined threshold.

In Fig.~\ref{fig:bl}, we illustrate an example of the baseline fitting using the arPLS method. The top panel shows the complete spectrum, while the middle and bottom panels show the zoom-in of the upper panel for a better view. We find that the arPLS method performs well in the Milky Way region ($\sim 1421$MHz in the bottom panel), the signal region ($\sim 1385$MHz in the middle panel), and the RFI-contaminated region ($1140-1300$ MHz). 

However, it is important to note that the iteration of the weight adjustment and application process in the arPLS method is time-intensive. For a rapid and efficient fitting of the baseline, an alternative method, arPoly, is provided, in which the Equation.~\ref{eq:bld-PLS} is replaced by a polynomial fitting. This method is significantly faster than arPLS, but is only suitable for simple baseline situations, such as fitting $T_{\mathrm{src}}^\mathrm{cont}$ after off-source subtraction.

One way to improve the quality of the baseline fitting is by stacking the spectra. Typically, the integration duration for recording a single spectrum does not exceed one second. In the case of wide-band data with a 7.6 kHz frequency resolution, noise levels during this short integration time are typically between 0.1 K and 0.3 K. These elevated noise levels can negatively impact the effectiveness of baseline removal algorithms, such as arPLS. To mitigate this issue, spectra recorded over several minutes are combined through stacking. This method presupposes the stability of the baseline over the duration, allowing the fitting of a common baseline to the combined spectra. Once this common baseline is determined and subtracted, a lower-order arPoly fit can be employed on each spectrum within the stack to adjust for any remaining baseline variations. Fig.~\ref{fig:waterfall} illustrates this process, where a comparison between panels C1 (C2) and D1 (D2) reveals the effective removal of the baseline.
 
\subsubsection{Standing wave fitting} \label{sec:sw}

The standing wave ${T}_{\mathrm{sw}}$ is induced by signals entering the receiving system with different paths and is widely observed in radio telescopes \citep[e.g.][]{Padman1977, Briggs1997, Peek2011}. For FAST, the primary standing wave arises from the reflection between the reflector and the receiver cabin \citep{Jiang2020}, where the path length difference of the signal is approximately 276 meters (approximately twice the distance between the reflector and the receiver cabin). This difference results in a delay of approximately 0.92 microseconds and gives rise to a standing wave with a period of approximately 1.09 MHz across the spectrum.

\HIFAST\ provides two methods for removing standing waves: sine-fitting and FFT-filter. Both approaches require a preliminary baseline removal. After isolating the standing wave, it is possible to subtract it from the spectra that include the baseline. This allows for the selective removal of only the standing wave, enabling off-source subtraction of the processed data.

The sine-fitting method utilizes the least-squares algorithm to fit the spectrum in the signal-free region using a sine function as follows:
\begin{equation}
   {T}_{\mathrm{sw}}(\nu) = a_0 + a_1\mathrm{sin}(2\pi f \nu + \phi), 
\end{equation}
where $a_0$, $a_1$, $f$, and $\phi$ are the standing wave's displacement, amplitude, frequency, and phase, respectively. The initial guess value of $f$ is set to $\sim 0.92$. With the baseline already subtracted, the signal-free regions are identifiable by selecting temperatures below a threshold, typically several times the root-mean-square (RMS) noise level of the spectra.

In the FFT-filter approach, the Fast Fourier Transform (FFT) is initially applied to the spectra, followed by the identification of modes associated with standing waves in Fourier space. The standing wave is then reconstructed through an inverse Fourier transformation of these modes. Illustrations of the FFT-filter technique's efficacy in removing standing waves are presented in panels C1 and C2 of Fig.~\ref{fig:waterfall}, which depict a waterfall plot and a 5-second integrated spectrum, respectively. Comprehensive details of the FFT-filter method, including criteria for selecting standing wave components in Fourier space and its comparison with alternative methods like running median, running mean, and sine-fitting, are elaborated in Paper III.

\subsubsection{RMS noise level}

\begin{figure*}
        \centering
	\includegraphics[width=2.\columnwidth]{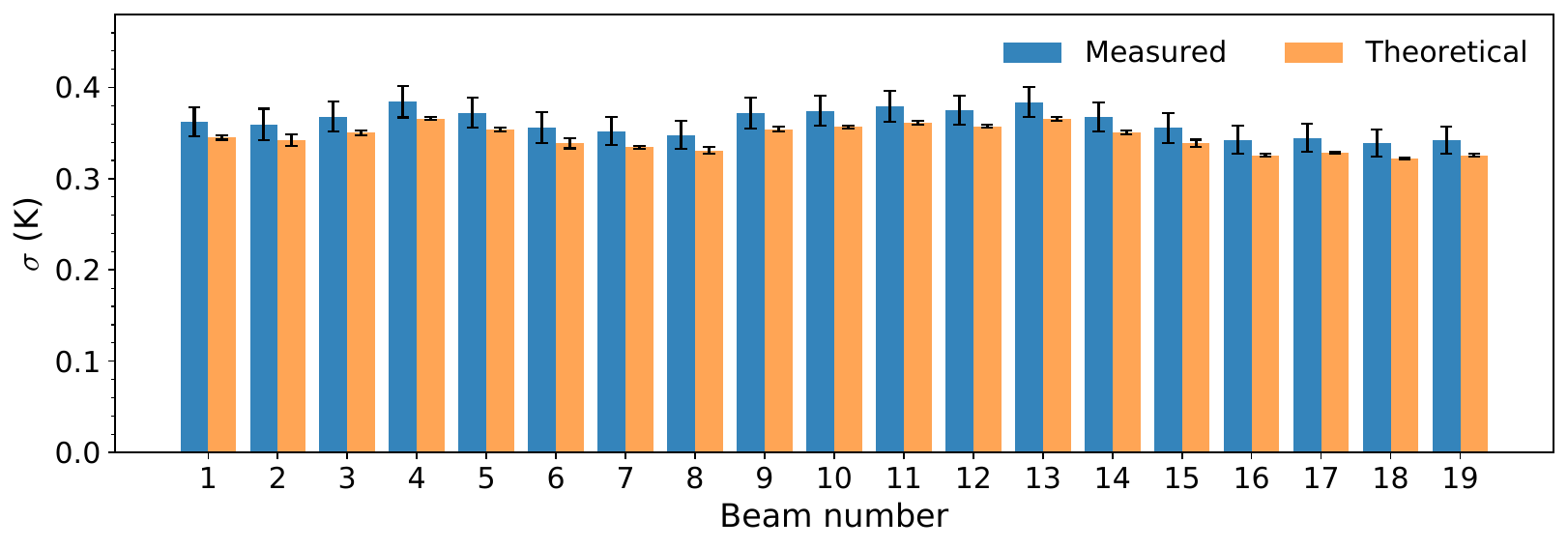}
	\caption{Comparison of RMS noise measured from spectra (blue) and the theoretical expectation (orange) for the XX polarization of each beam. The vertical bar represents the mean value of the 300 spectra for the XX polarization of each beam. The error bars represent the 1-sigma dispersion of the values.}
	\label{fig:rms}
\end{figure*}

In this subsection, we evaluate the calibration performance by comparing the measured root-mean-square (RMS) noise level with the theoretical expectations. Considering that the temperature and the RMS vary with frequency and time, we performed the test on 300 spectra with an integration time of 0.5 s and a frequency range ranging from 1415 to 1417 MHz.

For each spectrum, we estimate the theoretical RMS using the following equation:

\begin{equation}
\sigma_{\text{theoretical}} = \frac{T_{\text{input}}}{\sqrt{\Delta \nu \Delta t}}
\end{equation}

Here, $\Delta \nu$ represents the width of the spectral channel (7.6 kHz), and $\Delta t$ denotes the integration time (0.5 s). We selected $T_{\text{input}}$ as the temperature $T$ obtained from the temperature calibration as described in Section~\ref{sec:tcal}. To calculate $T_{\text{input}}$, we use the mean value of $T$ in the frequency range, which is typically approximately 20 K. It is important to note that the actual RMS was considered by measuring the spectra after removing the baseline and standing waves. Approximately 262 channels in the frequency range were used for the calculations.

Fig.~\ref{fig:rms} presents a comparison between the theoretical RMS (orange) and the actual measured RMS (blue) for the XX polarization of each beam. Our analysis revealed that the measured RMS values were slightly larger than the theoretical expectations. On average, the fractional difference, calculated by $(\sigma_{\text{measured}}-\sigma_{\text{theoretical}})/\sigma_{\text{theoretical}}$, is approximately 5\%, with a 1-sigma dispersion of approximately 4.5\% for each beam.

\subsection{Flux density calibration} \label{sec:flux}
Once the temperature of the line $T_{\mathrm{src}}^\mathrm{line}$ was determined, it was necessary to convert it into flux density by dividing it with the gain $G$ of FAST. The gain $G$ can vary depending on the telescope's condition during observation. Therefore, it is essential to observe a standard source with a precisely known flux as a calibrator observation before or after the target source observation. The gain $G$ is determined by calculating the ratio of the antenna temperature to the flux density of the calibrator. For detailed procedures on processing calibrator observation data, please refer to Paper II.

However, observing a calibrator every time may not always be feasible due to time constraints and the availability of calibrators. Alternatively, we can utilise a pre-measured gain value obtained from previous observations. A significant concern is the accuracy of these pre-measured gains. In Paper II, we examine the stability of gain $G$ over an extended period of observation. We found that using the pre-measured gain $G$ introduces an uncertainty of less than 3\% in general, although it can escalate to approximately 8\% in certain instances. We suggest that studies that need an accuracy better than 3\% in flux density should incorporate the observation of a calibrator together with their target sources.

\HIFAST\ provides two options for pre-determined gain values that are suitable for observations without a calibrator. The first option, as outlined in Paper II, has a high-frequency resolution of 5 MHz. The second option, as reported by \citet{Jiang2020}, is based on observations of the stable calibrator 3C286 using 19 FAST beams at varying zenith angles during August and December 2019. The frequency resolution in Paper II is finer than the 50 MHz resolution described in \citet{Jiang2020}.

\section{RFI flagging} \label{sec:rfi}

The signal of interest is inevitably contaminated by artificial radio emission, which is normally called Radio Frequency Interference (RFI). This noise is inevitable for large single-dish radio telescopes such as FAST. The RFI in FAST can be roughly divided into three types: harmonic RFI, wideband RFI, and narrowband RFI.

\begin{figure*}
        \centering
	\includegraphics[width=2\columnwidth]{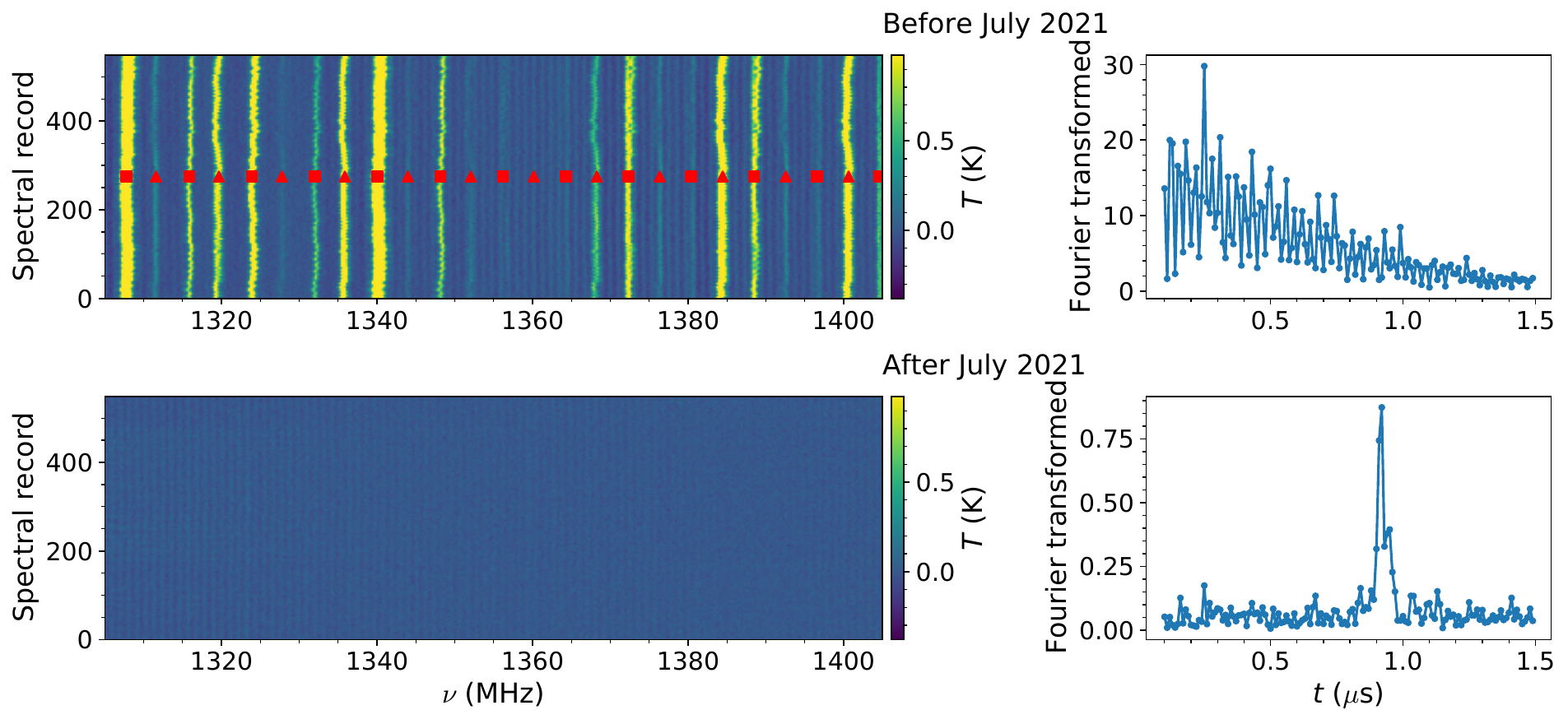}
	\caption{The 8.1 MHz harmonic RFI in the comparison of sample data observed before July 2021 (top panels) and after that (bottom panels). The left panels show the waterfall plot, i.e. temperature (coded by colour) as a function of frequency and the spectral record number for 500 spectra. The red square and triangle markers in the top left panel denote the position of two groups of 8.1 MHz RFIs, which disappear in the bottom left panel. The right panels show the Fourier space of 10-second integrated spectra selected from the corresponding left panel.}
	\label{fig:rfi-8mhz}
\end{figure*}

The harmonic RFI manifests as a sequence of bump signals with a coherent period in the frequency domain. Consequently, this kind of RFI disrupts the entire frequency range and effectively obstructs the channels of the target signal. We have detected several instances of such harmonic RFI. In this work, we provide a brief overview of an example with a period of approximately 8.1 MHz, which previously posed significant challenges to early \HI\ observations of FAST. Further information regarding other periods associated with this type of RFI can be found in Paper III. 

The 8.1 MHz harmonic RFI has a width of about 1 MHz and a period of approximately 8.1 MHz. In fact, there are two or three sets of harmonic RFI with an 8.1 MHz period in the spectra. Therefore, they can be easily identified as non-harmonic RFI \citep{Jiang2020, Zhang2021}. These sets of 8.1 MHz harmonic RFI spread over the entire bandwidth of the 19-beam L-band receiver, significantly impacting the observation of faint \HI\ sources. In the top panel of Figure \ref{fig:rfi-8mhz}, these RFIs are clearly visible in the waterfall plot (indicated by yellow and green vertical strips) and in the Fourier space of the 10-second integrated example spectrum (as high spikes). Specifically, the waterfall plot shows two groups of 8 MHz harmonic RFI, with the positions of the bumps marked by red square and triangle markers. Fortunately, the source of this RFI has been identified in the three sets of compressors in the cabin and was shielded in July 2021. The bottom panels show the results observed after July 2021. The interference disappeared in the waterfall plot, and only the 1 MHz standing wave component (0.92 $\mathrm{\mu s}$) remained.
 
\begin{figure*}
        \centering
	\includegraphics[width=2\columnwidth]{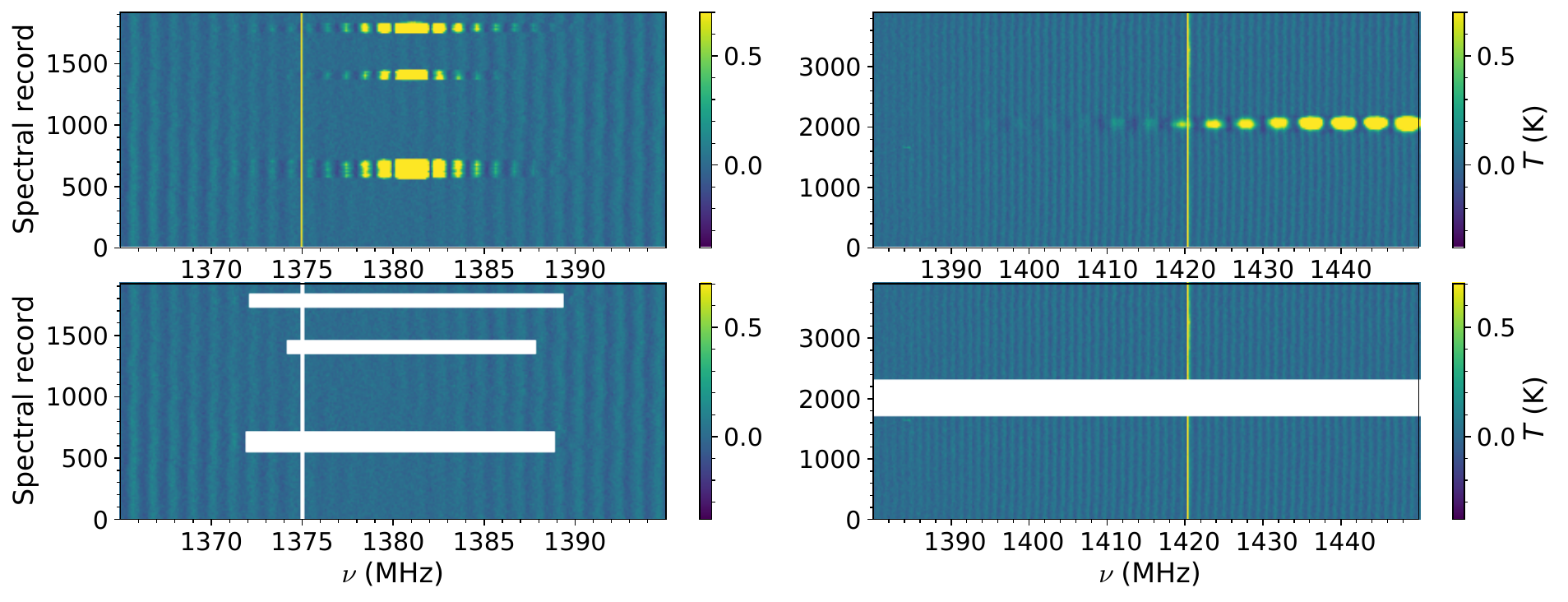}
	\caption{RFI flagging procedure for two different spectral frequency ranges. Each panel shows the spectra temperature (coded by colour) as a function of frequency and time. The upper panels show the original spectra, and the bottom panels show the corresponding RFI candidates are flagged as blank regions. Note that the regions near 1420 MHz are the Milky Way emissions, which are not flagged as desired.}
	\label{fig:rfi1}
\end{figure*}

The wideband RFI is caused by satellites or civil aviation in the sky, and primarily affects the spectra in the frequency range between 1155 and 1295 MHz. The probability of occurrence can reach 70\%, and even 99\% based on statistical analysis \citep[][based on $\sim 300$ hours of FAST observation data]{Zhang2021}. Observations in this frequency range should generally be avoided when using FAST. If necessary, a thorough check should be conducted before the observation by referring to \citet{Zhang2021}.

The narrowband RFI typically affects only one or two channels in spectra stored in the wide-band data with a frequency resolution of 7.6 kHz. An example can be found in the left panels of Fig.~\ref{fig:rfi1}, the region near 1375 MHz. The narrowband RFI was significantly reduced after April 2019 with the installation of electromagnetic shielding \citep{Jiang2020}, and there are only a few narrowband RFI left across the entire bandwidth (Fig.24 of \citet{Jiang2020}).

\HIFAST\ provides several automatic processes to mask the RFI considering that the RFI is different from the \HI\ signal in the polarization, time domain, and the possibility of accruing 19 beams. Usually, \HI\ emissions are not polarized and tend to last for a short time, not across multibeams. 

As an example, Fig.~\ref{fig:rfi1} displays a receipt to mask the RFI at 1381MHz related to GPS L3 emissions \citep[e.g.][]{Meyer2004,Haynes2018} in the left panels, and a wideband RFI larger than 1300 MHz possibly caused by geostationary satellites in the right panels. The upper panels illustrate the spectra temperature (coded by colour) in the frequency-time plane, and the bottom panels show the corresponding flagged RFI regions for RFI candidates as blank areas. The three regions centred at a frequency of 1380 MHz and located near the 550th, 1460th, and 1500th positions are examples of wideband RFI due to GPS L3 emissions. We identified and flagged them mainly based on their abrupt changes in time at the beginning and end of their occurrence. The right panels demonstrate a type of RFI that appeared sporadically over a wide frequency range larger than 50 MHz. The regions around 1420 MHz were kept because they were \HI\ emissions from the Milky Way.

Certainly, automatic processes cannot identify all RFI. \HIFAST\ offers a manual approach as a supplementary method to mask RFI. This involves selecting box regions to cover the RFI in the waterfall plots viewed in CARTA \citep{Comrie2021}.

\section{Doppler correction} \label{sec:corr_vel}

In FAST, observations occur in a topocentric mode, where the reference frame aligns with the telescope's geographical position. This frame differs from the standard in astrophysical research because it does not account for Earth's rotation, its orbital motion around the Sun, or the solar system's trajectory within the Milky Way. To rectify this and correct for the Doppler effect in the spectrum, a transformation to an alternate reference frame is necessary. \HIFAST\ offers two such frames: the Heliocentric frame, considering Earth's rotation and orbit around the Sun, and the Local Standard of Rest (LSR), which additionally incorporates the solar system's motion relative to nearby stars. 

Doppler correction varies depending on the time of observation, the pointing position in the sky, and the telescope's location. Although this variation might be minor in tracking mode, it is crucial in scanning mode, especially when spanning a wide sky area or conducting observations on different days. Therefore, in \HIFAST, the Doppler correction is applied individually to each spectrum. Subsequently, all spectra are interpolated onto a uniform frequency sampling grid.

\section{Imaging procedures}

\subsection{Stray radiation correction} \label{sec:sr}
The stray radiation arises from \HI\ emission entering the telescope receiver through the side lobes, rather than the main beam. In order to accurately determine the flux of the target sources, it is essential to take into account the effect of stray radiation. The measured temperatur $T_\mathrm{a}$ consists of two parts: one from the main beam (MB) and the other from the stray pattern (SP). Our algorithm does not attempt to resolve the temperature distribution in the main beam, but rather derives a beam-averaged brightness temperature ${T_{\mathrm{b}}}$ by solving the following equation:
\begin{align}
\label{eq:sp3}
{T_{\mathrm{b}}}(x, y)=&\frac{T_{\mathrm{a}}(x, y)}{\eta_{\mathrm{MB}}}\nonumber \\
&-\frac{1}{\eta_{\mathrm{MB}}} \int_{\mathrm{SP}} P\left(x-x^{\prime}, y-y^{\prime}\right) T_{\mathrm{b}}\left(x^{\prime}, y^{\prime}\right) \mathrm{d} x^{\prime} \mathrm{d} y^{\prime}.    
\end{align}
Here, $P(x, y)$ denotes the beam pattern and $\eta_{\mathrm{MB}}$ represents the main beam efficiency, defined as 
\begin{equation}
 \eta_{\mathrm{MB}} = \int_{\mathrm{MB}} P(x, y) \mathrm{d} x \mathrm{~d} y.
\end{equation}

This procedure is described in more detail in several publications, such as \citet[][]{Kalberla1980, Kalberla2005, Boothroyd2011, Winkel2016}. In \HIFAST, the correction is confined to near-field stray radiation and utilises the beam pattern, $P(x, y)$, from the observational results of Radio Galaxy PKS 0531+19 as described in Paper IV. We implement iterative processes, beginning with the observed $T_{\mathrm{a}}(x, y)$ as an initial approximation of $T_{\mathrm{b}}(x, y)$ on the right side of Equation~\ref{eq:sp3}, and then substituting the outcome into the subsequent iteration. This correction process is applied separately to each spectrum and each beam, but $T_{\mathrm{b}}(x, y)$ is updated simultaneously in the entire target range before the next iterative round.

This comprehensive procedure in \HIFAST\ can be found in Paper IV. In Paper IV, we also provide the beam pattern for each beam and a thorough examination of the stray radiation correction on the extended source M33 and point sources. After the stray radiation correction, the flux of point sources increases from 3\% to 6\% depending on the beam used, whereas the flux of some part of the extended source could be affected by more than $\pm 10\%$.

\subsection{Gridding map} \label{sec:cube}
To obtain different moment plots of the target source or region, the observed spectra should be properly assembled into a three-dimensional cube. The different moment plots can then be obtained by integrating the cube. The Nyquist sampling criteria for the number of spectra should be satisfied as the first principle. For FAST, the On-The-Fly (OTF) mapping and drift scanning observation modes have been meticulously designed to meet the Nyquist sampling criteria. This is achieved by rotating the 19-beam receiver at a specific angle, as detailed in Section 4.1 of \citet{Jiang2020}.

To assemble the individual spectra into cube data, normally one needs to assign the proper spectra into a regular grid in the Right Ascension (RA) – Declination (DEC) plane. This process is similar to creating a map of the sky, in which each grid point contains a single assembled spectrum. The initial step involves generating a World Coordinate System (WCS) header \citep[][]{Greisen2002}. The WCS is a standard framework in astronomy that associates a physical coordinate (here, RA and DEC) with each pixel in an image. This header corresponds to a regular grid based on the RA and DEC ranges of the input spectra. 

For each pixel in the grid, we consider the spectra located within a distance $r$ from the pixel, where $r$ is less than a predefined cutoff distance $r_\textrm{cut}$. These spectra contribute to the assembled spectrum in the pixel using a proper weighting method. This process ensures that only the relevant data are included for each grid point. The corrupted spectra are discarded to maintain the integrity of the data.

A weight $w_k$ is assigned to each remaining spectrum using a Gaussian or Bessel*Gaussian kernel, following the method in \citet{Mangum2007}. This weighting process takes into account the distance $r$ of each spectrum from the grid point and the beam size to ensure the appropriate scaling of the data.

The spectrum $S$ on the grid is then calculated using the following formula:

\begin{equation}
    S= \frac{1}{\sum w_k(r)}\sum_{k=1}^{n}w_k(r)S_k,
\end{equation}

where $S_k$ represents the remaining spectra. The resulting 3D RA–DEC–velocity or RA–DEC–frequency cube data are stored in a standard FITS file. For the integration of the cube at different moments along different dimensions, one could obtain the \HI\ density map, position–velocity map, velocity map, and velocity dispersion map, which are widely used to investigate the physical processes occurring in the target sources.

\section{Test Results}

This section presents the results of scientific tests that demonstrate the calibration and imaging performance of \HIFAST. First, we provide a brief introduction to the observation data used. Then, we discuss the results for the extended source (M33 galaxy) and point sources separately.

The observed sky covers the regions with $-5.2^\circ < \text{RA} < 27^\circ$ and $29.0^\circ < \text{DEC} < 36.5^\circ$. Within this region, M33, one of the galaxies closest to us, is prominently situated. Its position and the extensive observations previously conducted by the Arecibo telescope make it an exemplary target for our test on extended sources. At the same time, this specific celestial region overlaps significantly with the Arecibo Legacy Fast ALFA \citep[ALFALFA,][]{Haynes2011, Haynes2018}, carried out using the Arecibo telescope, making it equally suitable for the examination of point sources.

Drift scanning was utilised to observe these target regions. The 19-beam receiver array was rotated by $23.4^\circ$ during scanning to satisfy the Nyquist sampling criterion. To calibrate the temperature, a high noise diode ($\sim 10$ K) was injected for a duration of 2 seconds every 5 minutes. The quasar 3C48, which is close to M33, was observed on the same day as the flux calibrator. Each beam recorded the spectra at an exposure time of 0.5 second. The frequency resolution was approximately 7.63 kHz, corresponding to a velocity resolution of approximately 1.61 km $\rm s^{-1}$ at 1420 MHz.

\subsection{\HI\ map of M33 region}
\subsubsection{Data reduction}

The RAW data were processed using the \HIFAST\ modules described in this paper (see also Fig.~\ref{fig:flow} and Fig.~\ref{fig:waterfall}). Specifically, the system temperature was removed by combining the baseline and standing wave fitting. The spectral channels were shifted to the heliocentric frame. The pixel size was set to 1$^{\prime}$ during gridding, and spectra within a radius of 4.14$^{\prime}$ were considered with a weight of a Gaussian kernel with a sigma of 1.38$^{\prime}$. Eventually, we obtained a data cube with a velocity range of $-1500$ km $\rm s^{-1}$ to $500$ km $\rm s^{-1}$.

\subsubsection{Comparison with AGES} \label{sec:ages}
We compared our data cube with that of the Arecibo Galaxy Environment Survey \citep[AGES,][]{Keenan2016}. To maintain consistency, we used the same pixel size (i.e. 1$^{\prime}$) and grids in the RA-DEC plane as the data cube \footnote{\url{http://www.naic.edu/~ages/public_data.html}} from AGES. Since these two observations have different beam sizes, we convolved our data cube to match the beam size of the AGES. During this process, we assumed the beam shape to be a two-dimensional Gaussian function with a full width at half maximum (FWHM) equal to the beam size. Consequently, this assumption and the measurement of the beam size affected the comparisons related to the flux density. 

We binned the spectral channels by a factor of 3, resulting in a change in the velocity resolution from 1.61 km s$^{-1}$ to 4.83 km s$^{-1}$. This adjustment brings our resolution closer to the 5.2 km s$^{-1}$ resolution of AGES. Subsequently, we aligned our data cube with AGES by interpolating it onto the same frequency sampling grid. We subtracted a linear baseline along the RA direction for each channel in the cube data to adhere the treatment on \HI\ of Milky Way in AGES. To further minimize interference from the Milky Way, we computed the Moments in the velocity range of $-350$ to $-75$ km s$^{-1}$. Additionally, during the calculation, we used a mask thresholding at $5\sigma$ of the noise of the spectra during the calculation process to reduce potential data artifacts. Our analysis revealed a systematic velocity offset of approximately 4.75 km s$^{-1}$ between the FAST and AGES data, which falls within the velocity resolution of AGES. This offset, which likely arises from differences in velocity resolution, is discussed in detail in Appendix A. We have adjusted for this systematic offset in our comparative analysis to ensure accuracy.

\begin{figure*}
        \centering
	\includegraphics[width=2.\columnwidth]{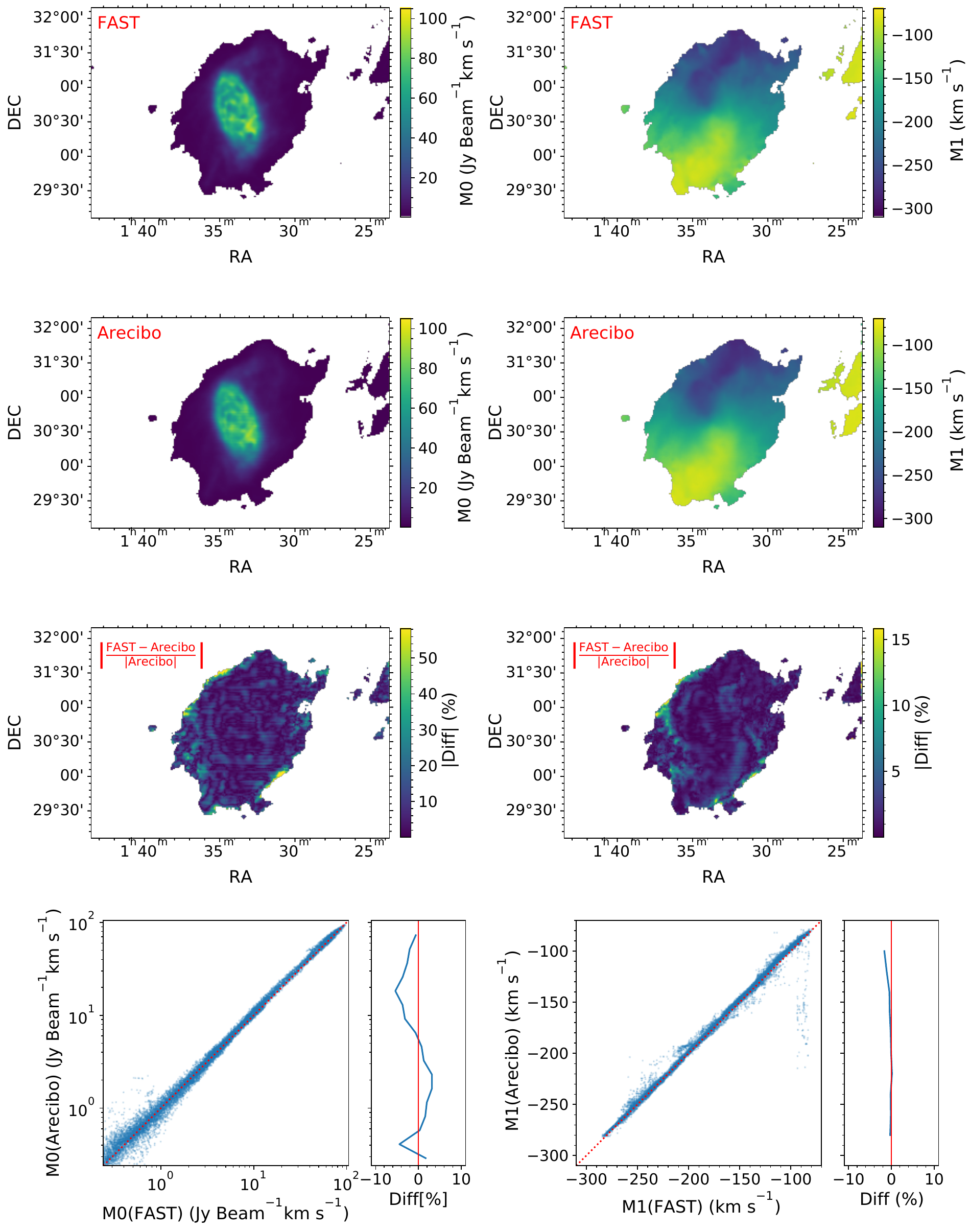}
	\caption{Comparisons of Moment-0 (M0, left panels) and Moment-1 (M1, right panels) of M33 \HI\ from FAST (this work) and Arecibo (AGES). The sequence includes FAST data maps (top row), Arecibo data maps (second row), and the absolute value of fractional difference between the two datasets (third row), calculated using the formula \((\text{FAST} - \text{Arecibo}) / | \text{Arecibo} | \times 100\%\). The fourth row contains scatter plots on the left, comparing pixel values from FAST and Arecibo, and plots on the right, displaying the median value of fractional difference as a function of the Arecibo values; the red lines in these plots denote the one-to-one correspondence between the two datasets.}
	\label{fig:m33-1}
\end{figure*}

In Figure~\ref{fig:m33-1}, we present a comparison of the integrated flux density (Moment-0) and the intensity-weighted velocities (Moment-1) in the left and right panels, respectively. The top and middle panels illustrate the \HI\ maps acquired from the FAST and Arecibo data, respectively. The third row displays a map showing the fractional difference between these observations, while a scatter plot is used for pixel-by-pixel comparisons.

The Moment-0 maps clearly demonstrate the structural similarities between the inner and outer regions of M33 in both observations. To quantify this agreement, a measure of the fractional difference is used, which is typically less than 10 percent. This scatter is likely attributable to systematic differences between the two telescopes, encompassing factors such as facility stability, receiver efficiency, and variations in post-processing of the data, among others. One notable source of this scatter is the dissimilar beam shapes of the two telescopes. FAST is equipped with 19 beam receivers, and apart from the central beam (beam 0), the shapes of the other beams are irregular, with slight differences in beam size \citep[refer to ][ and our Paper IV for more details]{Jiang2020}. We note that the relatively high fractional difference in certain pixels far away from M33 disk is due to the interference from the Milky Way and the velocity cut-off at $-75$ km s$^{-1}$. This will introduce a non-negligible error in the determination of the Moment-0 and subsequent Moments calculations.

The Moment-1 maps also exhibit a significant level of similarity in their structure, and most of the pixels in the fractional difference map deviate by less than 5 percent. However, it is worth mentioning that the median fractional difference is close to zero and does not show a strong correlation with the velocity values.

\begin{figure*}
        \centering
	\includegraphics[width=2.\columnwidth]{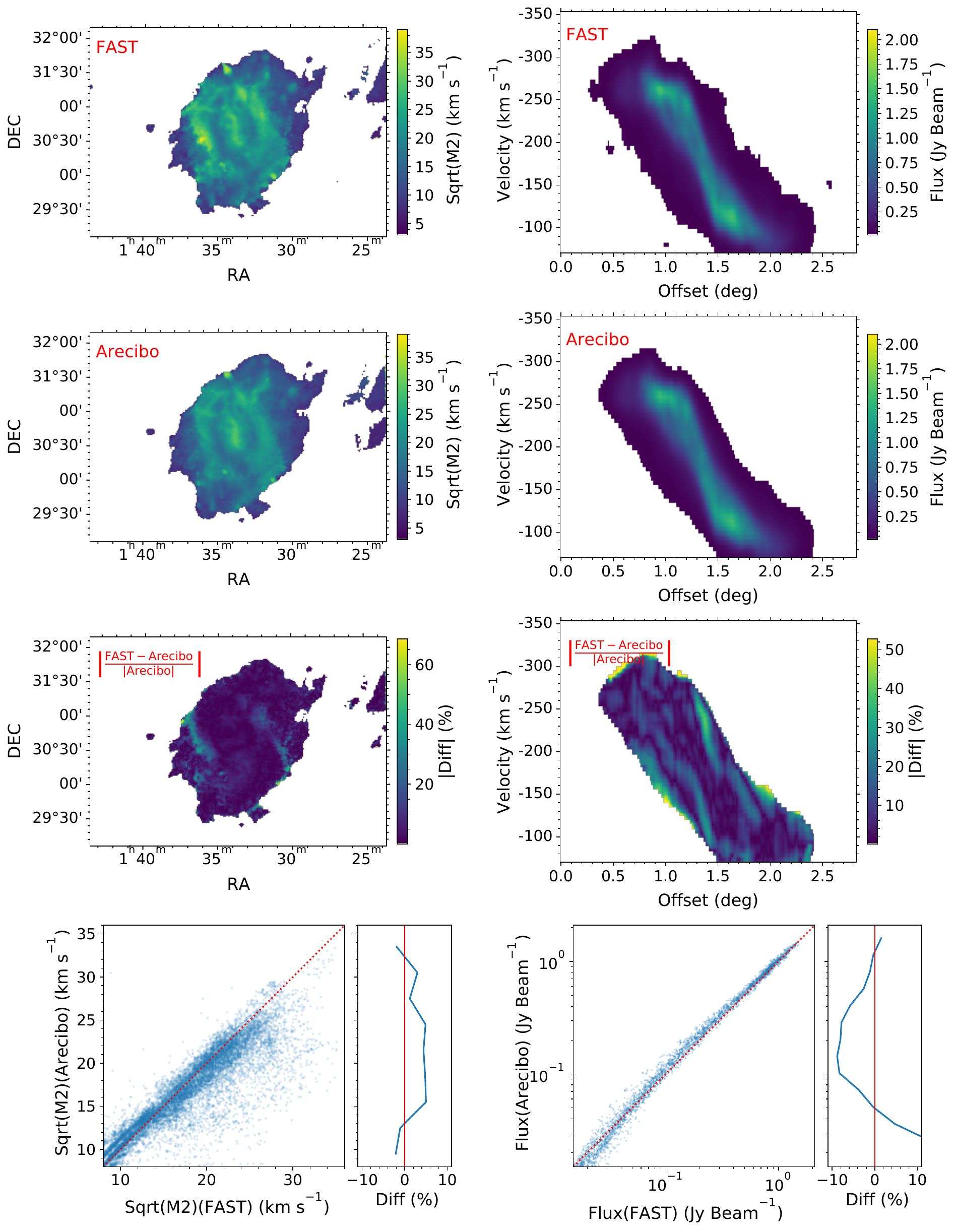}
	\caption{Similar to Fig.~\ref{fig:m33-1}, but for Moment-2 (left panels) and the position-velocity map (right panels). The position-velocity maps are extracted along the line of RA $=$ 01h33m51s. }
	\label{fig:m33-2}
\end{figure*}

Furthermore, we compared the weighted velocity dispersion (Moment-2, left panels) and position-velocity maps (right panels) of both datasets, as depicted in Fig.~\ref{fig:m33-2}. For Moment-2, the maps displayed a satisfactory level of agreement on structure. Pixel-by-pixel analysis shows that the median fractional difference is approximately 4 percent. Furthermore, the absolute difference is minor, approximately (0.6 km s$^{-1}$)$^2$. Analogous to the Moment-0 maps, the typical discrepancies in the flux density for each positive-velocity pixel were also below 10 percent.

In summary, considering the systematic differences between the two datasets, the agreement on Moments 0, 1, 2, and the position-velocity map is quite noteworthy.

\subsection{\HI\ Point sources}
\subsubsection{Data Reduction}
In this section, we analyse the FAST detected point sources within the frequency spectrum of 1300 to 1415 MHz. This particular frequency range has been selected due to its near absence of RFI and its overlap with Arecibo Legacy Fast ALFA \citep[ALFALFA,][]{Haynes2011, Haynes2018}. As the angular size of most galaxies in this frequency range is smaller than the FAST's beam size, we use the ``MinMed'' method to fit the off-source observation in system temperature removal (see Section~\ref{sec:ref}). The remaining processes were the same as those used to process the M33 \HI\ data cube. The sources were identified using SOFIA 2 software \citep{Serra2015, Westmeier2021}.

\subsubsection{Comparison with ALFALFA}
In this subsection, we compare the properties of the detected sources with the ALFALFA catalogue. The flux density $S_v$ from each source is measured by spatially integrating the spectra $s_v$ in the pixels as following \citep{Shostak1980}, given by
\begin{equation}
S_v=\frac{\sum_{x_0} \sum_{y_0} s_v(\Delta x, \Delta y)}{\sum_{x_0} \sum_{y_0} B(\Delta x, \Delta y)} \text {[Jy], }    
\end{equation}
where the numerator is the sum of $s_v$ over a selected solid angle covered by the region within the chosen isophotal fit \citep{Haynes2018} and the denominator is expressed as the sum of a centred, grid-sampled beam in the same region. The total integrated \HI\ line flux density $S_\mathrm{int}$ in Jy km s$^{-1}$ is then summed over all velocity ranges covered by the signal. Meanwhile, the signal-to-noise ratio $\mathbf{S/N}$ and the error of $S_\mathrm{int}$ are calculated using the same method used in \citet{Haynes2018}.

We cross-matched our sources with the ALFALFA based on their positions in the sky and the centre velocity. As \citet{Haynes2018} pointed out, the pipeline algorithm used for the ALFALFA catalogue may miss flux from very extended or highly asymmetric sources \citep{Haynes2018}. For a fair test, we limit the comparison to point sources with a characteristic size of the \HI\ disk smaller than 2$'$ (similar to \citet{Zuo2022}), which is less than the beam size of two telescopes. The characteristic size is estimated using the \HI\ size-mass relation in \citet{Wang2016}. We also limited the sample to $\mathbf{S/N} > 10$. This resulted in 221 sources matched from both datasets.

\begin{figure}[H]
        \centering
	\includegraphics[width=1.\columnwidth]{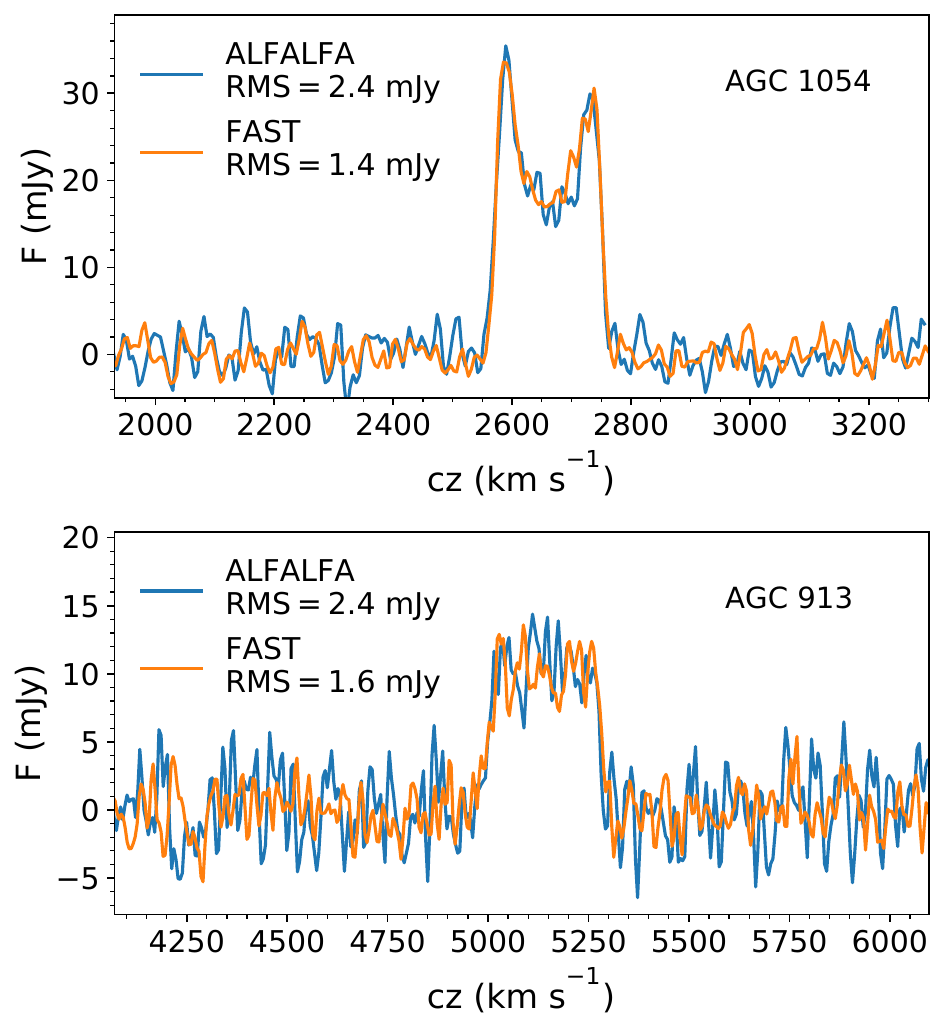}
	\caption{The spectra of two example galaxies detected in ALFALFA (blue lines) and FAST (orange lines). The  x-axis represents the optical velocity, corrected for the Heliocentric rest frame. Labels also include RMS noise of the spectra. The ALFALFA AGC number for each source is indicated in the top-right corner of the plot.}
	\label{fig:ps-s}
\end{figure}

First, two examples of spectra comparison are presented in Fig.~\ref{fig:ps-s}. The spectra' RMS noise are indicated on the labels. We notice that the spectra from FAST had lower RMS values with mean values of approximately $1.47$ and $2.24$ mJy for FAST and ALFALFA in 221 sources, respectively. Both observations were performed using the drifting scan mode. While the sampling spacing in DEC is around 1.05$^{\prime}$ \citep{Giovanelli2005} in ALFALFA's two-pass strategy, which is smaller than the spacing of approximately 1.14$^{\prime}$ in our FAST one-pass observation. This indicates a slightly longer integration observation time in ALFALFA. Given the shorter integration time and lower RMS noise, FAST demonstrated excellent performance in observing \HI.

\begin{figure}[H]
        \centering
	\includegraphics[width=1.\columnwidth]{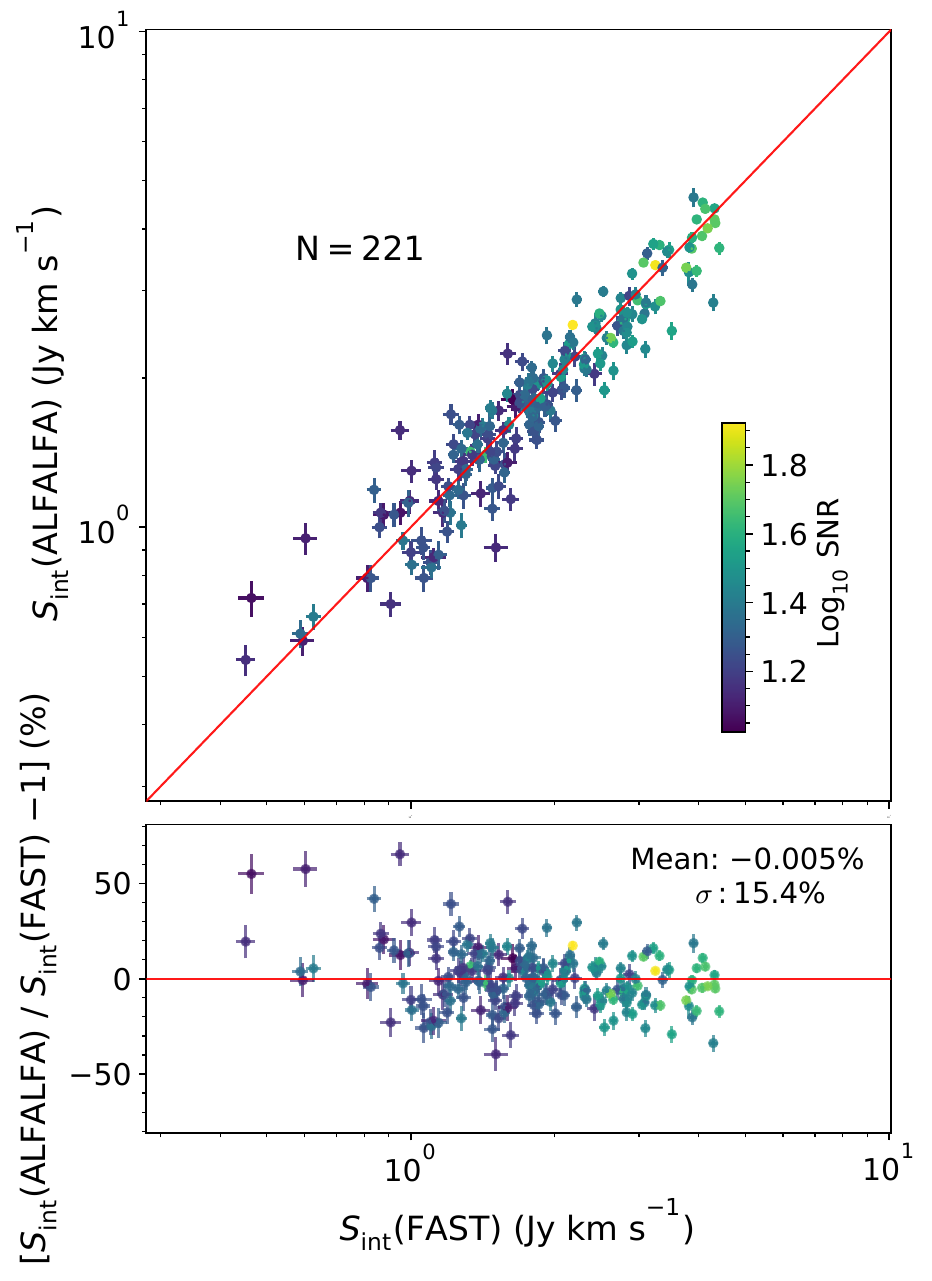}
	\caption{Comparison of integrated \HI\ line flux between FAST (this work) and the ALFALFA catalogue. The colour codes the signal-to-noise ratio of the source in FAST. Top panel: scatter plot of $S_\mathrm{int}$(FAST) versus $S_\mathrm{int}$(ALFALFA); Bottom panel: fractional difference as a function of $S_\mathrm{int}$(FAST). The red line in the top panel is a one-to-one relation and in the bottom panel the fractional difference equals 0.}
	\label{fig:ps-f}
\end{figure}

In Fig.~\ref{fig:ps-f}, we compare the integrated \HI\ line flux density $S_\mathrm{int}$ for 221 sources. The top panel displays the scatter plot of $S_\mathrm{int}$(FAST) versus $S_\mathrm{int}$(ALFALFA), while the bottom panel shows the fractional difference $S_\mathrm{int}$(ALFALFA)/$S_\mathrm{int}$(FAST)$-1$ as a function of $S_\mathrm{int}$(FAST). The data points are in close agreement with the equal value line (red line), with a mean fractional difference of approximately $-0.005$ per cent and a dispersion of 15.4 per cent. The scatter of the differences showed a slight dependence on flux: 16.1 per cent for $S_{\mathrm{int}}$(FAST) $<2.5$ Jy km s$^{-1}$ and 11.7 per cent for $S_{\mathrm{int}}$(FAST) $>2.5$ Jy km s$^{-1}$.

\begin{figure}[H]
        \centering
	\includegraphics[width=1.\columnwidth]{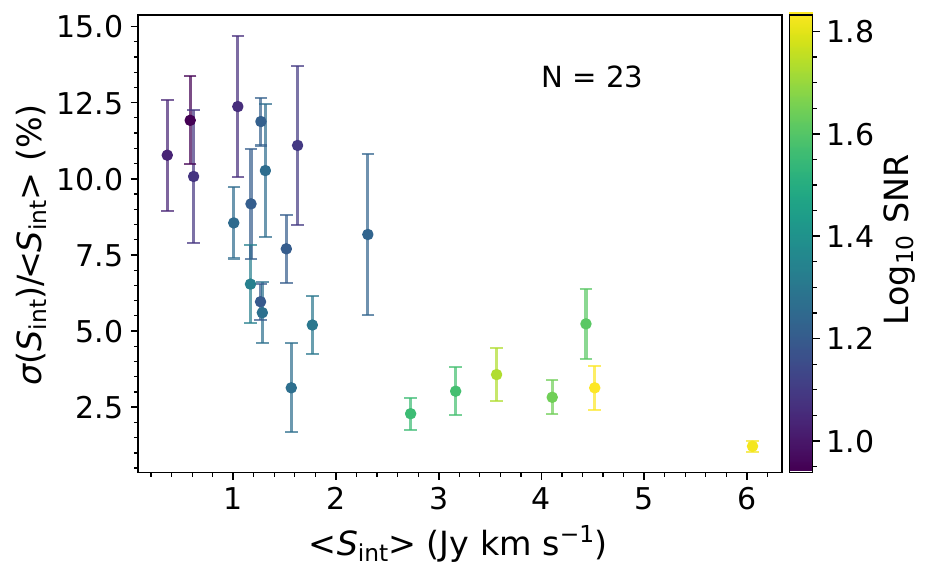}
	\caption{Multiple observation results for 23 sources. The x-axis and y-axis denote the mean and 1-sigma dispersion of integrated \HI\ line flux density measured in 7 times observations, respectively. The error bars are estimated by the bootstrap resampling method. The colour codes the mean signal-to-noise ratio.}
	\label{fig:ps-f-7r}
\end{figure}

Given that the data from ALFALFA and FAST were obtained from different facilities and processed through distinct pipelines, the scatter observed in the comparison between the two datasets could potentially be induced from various sources. These include differences in baseline complexity, uncertainties in beam shape, the multibeam synthesis (7 beams in ALFALFA and 19 beams in FAST), and uncertainties in source parameter estimation from the data cube. To estimate the inherent systematic error in FAST with \HIFAST\, we conducted a meticulous examination of multiple observations of the same sources. A 10 square-degree region of the sky was scanned seven times over a few months, leading to the identification of 23 sources.

In Fig.~\ref{fig:ps-f-7r}, we illustrate the 1-$\sigma$ dispersion of integrated \HI\ line flux density $S_\mathrm{int}$ measured in seven observations as a function of the mean value. The error for the dispersion is estimated using the bootstrap resampling method. For sources with $S_\mathrm{int}$ $>2.5$ Jy km s$^{-1}$, the dispersion is less than 5 per cent. However, for $S_\mathrm{int}$ $< 2.5$ Jy km s$^{-1}$, the dispersion increases to approximately 10 per cent as the uncertainty in the calibration procedure (particularly in system temperature removal operations) becomes more pronounced for fainter sources.

\section{Code Implementation and parallelization}
\HIFAST\ is written in Python programming language \citep{Python} , using the Python standard library and several third-party libraries: NumPy \citep{NumPy} and SciPy \citep{SciPy} for data structure and calculating; Astropy \citep{Astropy} for the calculation of RA-DEC and Doppler correction; Matplotlib \citep{Matplotlib} for diagnostic plotting; and h5py \citep{h5py} for intermediate data storage (a comprehensive list is available on our website). \HIFAST\ is an ongoing project, and its dependencies may change in the future.

Parallel calibration is a critical feature in \HIFAST\ because of the significant amount of data recorded by the 19-beam receiver of FAST. The data from the 19 beams were stored in separate files, and the majority of the procedures described previously were performed on each beam. This allowed \HIFAST\ to facilitate a streamlined simultaneous calibration. For certain time-consuming calibration processes, e.g., baseline fitting, stray radiation correction, and gridding, we employed the multiprocessing package to parallelize the calibration on each spectrum or the computation of the spectrum at each grid point.

\section{Discussion and Summary}

We have presented \HIFAST, a pipeline for processing \HI\ data obtained from FAST observations. This pipeline is compatible with tracking, drift scanning, and OTF mapping modes, as well as most of their variants. \HIFAST\ provides modular tasks to calibrate the raw data and create a data cube from the spectra. These tasks include temperature calibration with a noise diode as reference power input; system temperature removal using the ``MinMed'' method, baseline and standing wave fitting; flux density calibration by using a calibrator observed on the same day with the target or applying a predetermined gain function with zenith angle; an automated process to mask common RFI; Doppler correction of spectra from topocentric to Heliocentric or LSR reference frame; an iterative process to correct stray radiation; and interpolation of spectra to a regular grid of a data cube.

To assess the performance of \HIFAST, we used \HI\ data from the extended source M33 and point sources within the 1300 to 1415 MHz frequency range. After processing this RAW data with the pipeline, we produced an \HI\ image of M33 and a catalogue of point sources. Our comparison of the M33 data with the AGES results showed similar structures in the moment maps (0, 1, and 2), with a median fractional difference of less than 10 per cent. Additionally, we cross-matched our point sources with the ALFALFA survey catalog. In a common sample of 221 sources with a signal-to-noise ratio (S/N) greater than 10 from both datasets, the mean fractional difference in integrated flux density $S_{\mathrm{int}}$ was approximately 0.005 per cent, with a dispersion of 15.4 per cent. This dispersion varied with the value of $S_{\mathrm{int}}$: 16.1 per cent at $S_{\mathrm{int}}$(FAST) $<2.5$ Jy km s$^{-1}$ and 11.7 per cent at $S_{\mathrm{int}}$(FAST) $>2.5$ Jy km s$^{-1}$. Using multiple observations with FAST, we estimated the inherent systematic uncertainty of \HIFAST\ to be less than 5 per cent for sources with $S_{\mathrm{int}}>2.5$ Jy km s$^{-1}$ and about 10 per cent for those with $S_{\mathrm{int}}< 2.5$ Jy km s$^{-1}$.

\HIFAST\ has been successfully employed to process the FAST data in a number of studies, such as the FAST All Sky \HI\ survey \citep[FASHI,][]{Zhang2024}, M106 \citep{Zhu2021}, M51 \citep{Yu2023}, DDO168 \citep{Yu2023_2}, M94 \citep{Zhou2023}, and NGC4490/85 \citep{Liu2023}. While \HIFAST\ focuses mainly on processing the \HI\ data, its capabilities can be readily extended to handle OH line \citep{zhang2024OH} and ultra-wide band (UWB) data from FAST \citep{Zhang2023} in the future.


{\bf Acknowledgements.}
We would like to express our gratitude to Prof. Rendong Nan and his team for their groundbreaking efforts in establishing the FAST telescope. Additionally, we would like to acknowledge the FAST operation team for their invaluable assistance and support in the development of this pipeline. We thanks the referees for their constructive comments and suggestions. We also thank Martha P. Haynes for generously sharing ALFALFA spectra data and insightful discussions on source flux calculations, which significantly enhanced our study. We acknowledge the support of the China National Key Program for Science and Technology Research and Development of China (2022YFA1602901), the National Natural Science Foundation of China (Nos. 11988101, 11873051, 12125302, 12373011), the CAS Project for Young Scientists in Basic Research Grant (No. YSBR-062), and the K.C. Wong Education Foundation, and the science research grants from the
China Manned Space Project. Y Jing acknowledges support from the Cultivation Project for FAST Scientific Payoff and Research Achievement of CAMS-CAS. 


{\bf InterestConflict.}
The authors declare that they have no conflict of interest.


\bibliographystyle{aasjournal}
\bibliography{main} 

\begin{thebibliography}{}
\expandafter\ifx\csname natexlab\endcsname\relax\def\natexlab#1{#1}\fi
\providecommand{\url}[1]{\href{#1}{#1}}
\providecommand{\dodoi}[1]{doi:~\href{http://doi.org/#1}{\nolinkurl{#1}}}
\providecommand{\doeprint}[1]{\href{http://ascl.net/#1}{\nolinkurl{http://ascl.net/#1}}}
\providecommand{\doarXiv}[1]{\href{https://arxiv.org/abs/#1}{\nolinkurl{https://arxiv.org/abs/#1}}}

\bibitem[{{Astropy Collaboration} {et~al.}(2022){Astropy Collaboration},
  {Price-Whelan}, {Lim}, {Earl}, {Starkman}, {Bradley}, {Shupe}, {Patil},
  {Corrales}, {Brasseur}, {N{\"o}the}, {Donath}, {Tollerud}, {Morris},
  {Ginsburg}, {Vaher}, {Weaver}, {Tocknell}, {Jamieson}, {van Kerkwijk},
  {Robitaille}, {Merry}, {Bachetti}, {G{\"u}nther}, {Aldcroft},
  {Alvarado-Montes}, {Archibald}, {B{\'o}di}, {Bapat}, {Barentsen},
  {Baz{\'a}n}, {Biswas}, {Boquien}, {Burke}, {Cara}, {Cara}, {Conroy},
  {Conseil}, {Craig}, {Cross}, {Cruz}, {D'Eugenio}, {Dencheva}, {Devillepoix},
  {Dietrich}, {Eigenbrot}, {Erben}, {Ferreira}, {Foreman-Mackey}, {Fox},
  {Freij}, {Garg}, {Geda}, {Glattly}, {Gondhalekar}, {Gordon}, {Grant},
  {Greenfield}, {Groener}, {Guest}, {Gurovich}, {Handberg}, {Hart},
  {Hatfield-Dodds}, {Homeier}, {Hosseinzadeh}, {Jenness}, {Jones}, {Joseph},
  {Kalmbach}, {Karamehmetoglu}, {Ka{\l}uszy{\'n}ski}, {Kelley}, {Kern},
  {Kerzendorf}, {Koch}, {Kulumani}, {Lee}, {Ly}, {Ma}, {MacBride}, {Maljaars},
  {Muna}, {Murphy}, {Norman}, {O'Steen}, {Oman}, {Pacifici}, {Pascual},
  {Pascual-Granado}, {Patil}, {Perren}, {Pickering}, {Rastogi}, {Roulston},
  {Ryan}, {Rykoff}, {Sabater}, {Sakurikar}, {Salgado}, {Sanghi}, {Saunders},
  {Savchenko}, {Schwardt}, {Seifert-Eckert}, {Shih}, {Jain}, {Shukla}, {Sick},
  {Simpson}, {Singanamalla}, {Singer}, {Singhal}, {Sinha}, {Sip{\H{o}}cz},
  {Spitler}, {Stansby}, {Streicher}, {{\v{S}}umak}, {Swinbank}, {Taranu},
  {Tewary}, {Tremblay}, {de Val-Borro}, {Van Kooten}, {Vasovi{\'c}}, {Verma},
  {de Miranda Cardoso}, {Williams}, {Wilson}, {Winkel}, {Wood-Vasey}, {Xue},
  {Yoachim}, {Zhang}, {Zonca}, \& {Astropy Project Contributors}}]{Astropy}
{Astropy Collaboration}, {Price-Whelan}, A.~M., {Lim}, P.~L., {et~al.} 2022,
  \apj, 935, 167, \dodoi{10.3847/1538-4357/ac7c74}

\bibitem[{Baek {et~al.}(2015)Baek, Park, Ahn, \& Choo}]{Baek2015}
Baek, S.-J., Park, A., Ahn, Y.-J., \& Choo, J. 2015, Analyst, 140, 250,
  \dodoi{10.1039/C4AN01061B}

\bibitem[{{Boothroyd} {et~al.}(2011){Boothroyd}, {Blagrave}, {Lockman},
  {Martin}, {Pinheiro Gon{\c{c}}alves}, \& {Srikanth}}]{Boothroyd2011}
{Boothroyd}, A.~I., {Blagrave}, K., {Lockman}, F.~J., {et~al.} 2011, \aap, 536,
  A81, \dodoi{10.1051/0004-6361/201117656}

\bibitem[{{Briggs} {et~al.}(1997){Briggs}, {Sorar}, {Kraan-Korteweg}, \& {van
  Driel}}]{Briggs1997}
{Briggs}, F.~H., {Sorar}, E., {Kraan-Korteweg}, R.~C., \& {van Driel}, W. 1997,
  \pasa, 14, 37, \dodoi{10.1071/AS97037}

\bibitem[{Collette(2013)}]{h5py}
Collette, A. 2013, Python and HDF5 (O'Reilly)

\bibitem[{Comrie {et~al.}(2021)Comrie, Wang, Hsu, Moraghan, Harris, Pang,
  Pińska, Chiang, Chang, Hwang, Jan, Lin, \& Simmonds}]{Comrie2021}
Comrie, A., Wang, K.-S., Hsu, S.-C., {et~al.} 2021, {CARTA: The Cube Analysis
  and Rendering Tool for Astronomy}, 2.0.0,  Zenodo,
  \dodoi{10.5281/zenodo.4905459}

\bibitem[{{Davies} {et~al.}(2011){Davies}, {Auld}, {Burns}, {Minchin},
  {Momjian}, {Schneider}, {Smith}, {Taylor}, \& {van Driel}}]{Davies2011}
{Davies}, J.~I., {Auld}, R., {Burns}, L., {et~al.} 2011, \mnras, 415, 1883,
  \dodoi{10.1111/j.1365-2966.2011.18833.x}

\bibitem[{Dunning {et~al.}(2017)Dunning, Bowen, Castillo, Chung, Doherty,
  George, Hayman, Jeganathan, Kanoniuk, Mackay, Reilly, Roush, Smart, Shaw,
  Smith, Tzioumis, \& Venables}]{Dunning2017}
Dunning, A., Bowen, M., Castillo, S., {et~al.} 2017, in 2017 XXXIInd General
  Assembly and Scientific Symposium of the International Union of Radio Science
  (URSI GASS), 1--4, \dodoi{10.23919/URSIGASS.2017.8105012}

\bibitem[{Eilers(2003)}]{Eilers2003}
Eilers, P. H.~C. 2003, Analytical Chemistry, 75, 3631,
  \dodoi{10.1021/ac034173t}

\bibitem[{{Giovanelli} {et~al.}(2005){Giovanelli}, {Haynes}, {Kent},
  {Perillat}, {Catinella}, {Hoffman}, {Momjian}, {Rosenberg}, {Saintonge},
  {Spekkens}, {Stierwalt}, {Brosch}, {Masters}, {Springob}, {Karachentsev},
  {Karachentseva}, {Koopmann}, {Muller}, {van Driel}, \& {van
  Zee}}]{Giovanelli2005}
{Giovanelli}, R., {Haynes}, M.~P., {Kent}, B.~R., {et~al.} 2005, \aj, 130,
  2613, \dodoi{10.1086/497432}

\bibitem[{{Greisen} \& {Calabretta}(2002)}]{Greisen2002}
{Greisen}, E.~W., \& {Calabretta}, M.~R. 2002, \aap, 395, 1061,
  \dodoi{10.1051/0004-6361:20021326}

\bibitem[{Harris {et~al.}(2020)Harris, Millman, van~der Walt, Gommers,
  Virtanen, Cournapeau, Wieser, Taylor, Berg, Smith, Kern, Picus, Hoyer, van
  Kerkwijk, Brett, Haldane, del R{\'{i}}o, Wiebe, Peterson,
  G{\'{e}}rard-Marchant, Sheppard, Reddy, Weckesser, Abbasi, Gohlke, \&
  Oliphant}]{NumPy}
Harris, C.~R., Millman, K.~J., van~der Walt, S.~J., {et~al.} 2020, Nature, 585,
  357, \dodoi{10.1038/s41586-020-2649-2}

\bibitem[{{Haynes} {et~al.}(2011){Haynes}, {Giovanelli}, {Martin}, {Hess},
  {Saintonge}, {Adams}, {Hallenbeck}, {Hoffman}, {Huang}, {Kent}, {Koopmann},
  {Papastergis}, {Stierwalt}, {Balonek}, {Craig}, {Higdon}, {Kornreich},
  {Miller}, {O'Donoghue}, {Olowin}, {Rosenberg}, {Spekkens}, {Troischt}, \&
  {Wilcots}}]{Haynes2011}
{Haynes}, M.~P., {Giovanelli}, R., {Martin}, A.~M., {et~al.} 2011, \aj, 142,
  170, \dodoi{10.1088/0004-6256/142/5/170}

\bibitem[{{Haynes} {et~al.}(2018){Haynes}, {Giovanelli}, {Kent}, {Adams},
  {Balonek}, {Craig}, {Fertig}, {Finn}, {Giovanardi}, {Hallenbeck}, {Hess},
  {Hoffman}, {Huang}, {Jones}, {Koopmann}, {Kornreich}, {Leisman}, {Miller},
  {Moorman}, {O'Connor}, {O'Donoghue}, {Papastergis}, {Troischt}, {Stark}, \&
  {Xiao}}]{Haynes2018}
{Haynes}, M.~P., {Giovanelli}, R., {Kent}, B.~R., {et~al.} 2018, \apj, 861, 49,
  \dodoi{10.3847/1538-4357/aac956}

\bibitem[{Hunter(2007)}]{Matplotlib}
Hunter, J.~D. 2007, Computing in Science \& Engineering, 9, 90,
  \dodoi{10.1109/MCSE.2007.55}

\bibitem[{Jeganathan {et~al.}(2017)Jeganathan, Dunning, Chengjin, Haiyan,
  Chung, Mackay, Doherty, Kanoniuk, Reilly, Roush, Shaw, Severs, Bowen, \&
  Hayman}]{Jeganathan2017}
Jeganathan, K., Dunning, A., Chengjin, J., {et~al.} 2017, in 2017 IEEE Asia
  Pacific Microwave Conference (APMC), 244--247,
  \dodoi{10.1109/APMC.2017.8251424}

\bibitem[{{Jiang} {et~al.}(2019){Jiang}, {Yue}, {Gan}, {Yao}, {Li}, {Pan},
  {Sun}, {Yu}, {Liu}, {Tang}, {Qian}, {Lu}, {Yan}, {Peng}, {Zhang}, {Wang},
  {Li}, \& {Li}}]{Jiang2019}
{Jiang}, P., {Yue}, Y., {Gan}, H., {et~al.} 2019, Science China Physics,
  Mechanics, and Astronomy, 62, 959502, \dodoi{10.1007/s11433-018-9376-1}

\bibitem[{{Jiang} {et~al.}(2020){Jiang}, {Tang}, {Hou}, {Liu}, {Kr{\v{c}}o},
  {Qian}, {Sun}, {Ching}, {Liu}, {Duan}, {Yue}, {Gan}, {Yao}, {Li}, {Pan},
  {Yu}, {Liu}, {Li}, {Peng}, {Yan}, \& {FAST Collaboration}}]{Jiang2020}
{Jiang}, P., {Tang}, N.-Y., {Hou}, L.-G., {et~al.} 2020, Research in Astronomy
  and Astrophysics, 20, 064, \dodoi{10.1088/1674-4527/20/5/64}

\bibitem[{{Kalberla} {et~al.}(2005){Kalberla}, {Burton}, {Hartmann}, {Arnal},
  {Bajaja}, {Morras}, \& {P{\"o}ppel}}]{Kalberla2005}
{Kalberla}, P.~M.~W., {Burton}, W.~B., {Hartmann}, D., {et~al.} 2005, \aap,
  440, 775, \dodoi{10.1051/0004-6361:20041864}

\bibitem[{{Kalberla} {et~al.}(1980){Kalberla}, {Mebold}, \&
  {Reich}}]{Kalberla1980}
{Kalberla}, P.~M.~W., {Mebold}, U., \& {Reich}, W. 1980, \aap, 82, 275

\bibitem[{{Keenan} {et~al.}(2016){Keenan}, {Davies}, {Taylor}, \&
  {Minchin}}]{Keenan2016}
{Keenan}, O.~C., {Davies}, J.~I., {Taylor}, R., \& {Minchin}, R.~F. 2016,
  \mnras, 456, 951, \dodoi{10.1093/mnras/stv2684}

\bibitem[{{Liu} {et~al.}(2023){Liu}, {Zhu}, {Yu}, {Ai}, {Jiang}, {Liu}, \&
  {Yuan}}]{Liu2023}
{Liu}, Y., {Zhu}, M., {Yu}, H., {et~al.} 2023, \mnras, 523, 3905,
  \dodoi{10.1093/mnras/stad1281}

\bibitem[{{Mangum} {et~al.}(2007){Mangum}, {Emerson}, \&
  {Greisen}}]{Mangum2007}
{Mangum}, J.~G., {Emerson}, D.~T., \& {Greisen}, E.~W. 2007, \aap, 474, 679,
  \dodoi{10.1051/0004-6361:20077811}

\bibitem[{{Meyer} {et~al.}(2004){Meyer}, {Zwaan}, {Webster}, {Staveley-Smith},
  {Ryan-Weber}, {Drinkwater}, {Barnes}, {Howlett}, {Kilborn}, {Stevens},
  {Waugh}, {Pierce}, {Bhathal}, {de Blok}, {Disney}, {Ekers}, {Freeman},
  {Garcia}, {Gibson}, {Harnett}, {Henning}, {Jerjen}, {Kesteven}, {Knezek},
  {Koribalski}, {Mader}, {Marquarding}, {Minchin}, {O'Brien}, {Oosterloo},
  {Price}, {Putman}, {Ryder}, {Sadler}, {Stewart}, {Stootman}, \&
  {Wright}}]{Meyer2004}
{Meyer}, M.~J., {Zwaan}, M.~A., {Webster}, R.~L., {et~al.} 2004, \mnras, 350,
  1195, \dodoi{10.1111/j.1365-2966.2004.07710.x}

\bibitem[{{Nan} {et~al.}(2011){Nan}, {Li}, {Jin}, {Wang}, {Zhu}, {Zhu},
  {Zhang}, {Yue}, \& {Qian}}]{Nan2011}
{Nan}, R., {Li}, D., {Jin}, C., {et~al.} 2011, International Journal of Modern
  Physics D, 20, 989, \dodoi{10.1142/S0218271811019335}

\bibitem[{{O'Neil}(2002)}]{Neil2002}
{O'Neil}, K. 2002, in Astronomical Society of the Pacific Conference Series,
  Vol. 278, Single-Dish Radio Astronomy: Techniques and Applications, ed.
  S.~{Stanimirovic}, D.~{Altschuler}, P.~{Goldsmith}, \& C.~{Salter}, 293--311.
\newblock \doarXiv{astro-ph/0203001}

\bibitem[{{Padman}(1977)}]{Padman1977}
{Padman}, R. 1977, \pasa, 3, 111, \dodoi{10.1017/S132335800001496X}

\bibitem[{{Peek} {et~al.}(2011){Peek}, {Heiles}, {Douglas}, {Lee}, {Grcevich},
  {Stanimirovi{\'c}}, {Putman}, {Korpela}, {Gibson}, {Begum}, {Saul},
  {Robishaw}, \& {Kr{\v{c}}o}}]{Peek2011}
{Peek}, J.~E.~G., {Heiles}, C., {Douglas}, K.~A., {et~al.} 2011, \apjs, 194,
  20, \dodoi{10.1088/0067-0049/194/2/20}

\bibitem[{{Peek} {et~al.}(2018){Peek}, {Babler}, {Zheng}, {Clark}, {Douglas},
  {Korpela}, {Putman}, {Stanimirovi{\'c}}, {Gibson}, \& {Heiles}}]{Peek2018}
{Peek}, J.~E.~G., {Babler}, B.~L., {Zheng}, Y., {et~al.} 2018, \apjs, 234, 2,
  \dodoi{10.3847/1538-4365/aa91d3}

\bibitem[{{Putman} {et~al.}(2003){Putman}, {Staveley-Smith}, {Freeman},
  {Gibson}, \& {Barnes}}]{Putman2003}
{Putman}, M.~E., {Staveley-Smith}, L., {Freeman}, K.~C., {Gibson}, B.~K., \&
  {Barnes}, D.~G. 2003, \apj, 586, 170, \dodoi{10.1086/344477}

\bibitem[{{Putman} {et~al.}(2002){Putman}, {de Heij}, {Staveley-Smith},
  {Braun}, {Freeman}, {Gibson}, {Burton}, {Barnes}, {Banks}, {Bhathal}, {de
  Blok}, {Boyce}, {Disney}, {Drinkwater}, {Ekers}, {Henning}, {Jerjen},
  {Kilborn}, {Knezek}, {Koribalski}, {Malin}, {Marquarding}, {Minchin},
  {Mould}, {Oosterloo}, {Price}, {Ryder}, {Sadler}, {Stewart}, {Stootman},
  {Webster}, \& {Wright}}]{Putman2002}
{Putman}, M.~E., {de Heij}, V., {Staveley-Smith}, L., {et~al.} 2002, \aj, 123,
  873, \dodoi{10.1086/338088}

\bibitem[{{Qian} {et~al.}(2020){Qian}, {Yao}, {Sun}, {Xu}, {Pan}, \&
  {Jiang}}]{Qian2020}
{Qian}, L., {Yao}, R., {Sun}, J., {et~al.} 2020, The Innovation, 1, 100053,
  \dodoi{10.1016/j.xinn.2020.100053}

\bibitem[{{Serra} {et~al.}(2015){Serra}, {Westmeier}, {Giese}, {Jurek},
  {Fl{\"o}er}, {Popping}, {Winkel}, {van der Hulst}, {Meyer}, {Koribalski},
  {Staveley-Smith}, \& {Courtois}}]{Serra2015}
{Serra}, P., {Westmeier}, T., {Giese}, N., {et~al.} 2015, \mnras, 448, 1922,
  \dodoi{10.1093/mnras/stv079}

\bibitem[{{Shostak} \& {Allen}(1980)}]{Shostak1980}
{Shostak}, G.~S., \& {Allen}, R.~J. 1980, \aap, 81, 167

\bibitem[{{Taylor} {et~al.}(2014){Taylor}, {Minchin}, {Herbst}, {Davies},
  {Rodriguez}, \& {Vazquez}}]{Taylor2014}
{Taylor}, R., {Minchin}, R.~F., {Herbst}, H., {et~al.} 2014, \mnras, 443, 2634,
  \dodoi{10.1093/mnras/stu1305}

\bibitem[{Van~Rossum \& Drake(2009)}]{Python}
Van~Rossum, G., \& Drake, F.~L. 2009, Python 3 Reference Manual (Scotts Valley,
  CA: CreateSpace)

\bibitem[{Virtanen {et~al.}(2020)Virtanen, Gommers, Oliphant, Haberland, Reddy,
  Cournapeau, Burovski, Peterson, Weckesser, Bright, {van der Walt}, Brett,
  Wilson, Millman, Mayorov, Nelson, Jones, Kern, Larson, Carey, Polat, Feng,
  Moore, {VanderPlas}, Laxalde, Perktold, Cimrman, Henriksen, Quintero, Harris,
  Archibald, Ribeiro, Pedregosa, {van Mulbregt}, \& {SciPy 1.0
  Contributors}}]{SciPy}
Virtanen, P., Gommers, R., Oliphant, T.~E., {et~al.} 2020, Nature Methods, 17,
  261, \dodoi{10.1038/s41592-019-0686-2}

\bibitem[{{Wang} {et~al.}(2016){Wang}, {Koribalski}, {Serra}, {van der Hulst},
  {Roychowdhury}, {Kamphuis}, \& {Chengalur}}]{Wang2016}
{Wang}, J., {Koribalski}, B.~S., {Serra}, P., {et~al.} 2016, \mnras, 460, 2143,
  \dodoi{10.1093/mnras/stw1099}

\bibitem[{{Westmeier} {et~al.}(2021){Westmeier}, {Kitaeff}, {Pallot}, {Serra},
  {van der Hulst}, {Jurek}, {Elagali}, {For}, {Kleiner}, {Koribalski},
  {Lee-Waddell}, {Mould}, {Reynolds}, {Rhee}, \&
  {Staveley-Smith}}]{Westmeier2021}
{Westmeier}, T., {Kitaeff}, S., {Pallot}, D., {et~al.} 2021, \mnras, 506, 3962,
  \dodoi{10.1093/mnras/stab1881}

\bibitem[{{Winkel} {et~al.}(2016){Winkel}, {Kerp}, {Fl{\"o}er}, {Kalberla},
  {Ben Bekhti}, {Keller}, \& {Lenz}}]{Winkel2016}
{Winkel}, B., {Kerp}, J., {Fl{\"o}er}, L., {et~al.} 2016, \aap, 585, A41,
  \dodoi{10.1051/0004-6361/201527007}

\bibitem[{{Winkel} {et~al.}(2012){Winkel}, {Kraus}, \& {Bach}}]{Winkel2012}
{Winkel}, B., {Kraus}, A., \& {Bach}, U. 2012, \aap, 540, A140,
  \dodoi{10.1051/0004-6361/201118092}

\bibitem[{{Wolfe} {et~al.}(2015){Wolfe}, {Pisano}, \& J.}]{Wolfe2015}
{Wolfe}, S.~A., {Pisano}, D.~J.~L., \& J., F. 2015, GBT Memo, 289

\bibitem[{{Yin} {et~al.}(2023){Yin}, {Jiang}, \& {Rao}}]{Yin2023}
{Yin}, J., {Jiang}, P., \& {Rao}, R. 2023, SCIENCE CHINA Physics, Mechanics \&
  Astronomy, 66, 239513, \dodoi{10.1007/s11433-022-1997-8}

\bibitem[{{Yu} {et~al.}(2023{\natexlab{a}}){Yu}, {Zhu}, {Xu}, {Ai}, {Jiang}, \&
  {Yang}}]{Yu2023}
{Yu}, H., {Zhu}, M., {Xu}, J.-L., {et~al.} 2023{\natexlab{a}}, \mnras, 521,
  2719, \dodoi{10.1093/mnras/stad436}

\bibitem[{{Yu} {et~al.}(2023{\natexlab{b}}){Yu}, {Zhu}, {Xu}, {Zhang}, {Liu},
  {Jiang}, \& {Wang}}]{Yu2023_2}
{Yu}, N.-P., {Zhu}, M., {Xu}, J.-L., {et~al.} 2023{\natexlab{b}}, \mnras, 521,
  737, \dodoi{10.1093/mnras/stad561}

\bibitem[{{Zhang} {et~al.}(2024{\natexlab{a}}){Zhang}, {Cheng}, {Zhu}, \&
  {Jiang}}]{zhang2024OH}
{Zhang}, C.-P., {Cheng}, C., {Zhu}, M., \& {Jiang}, P. 2024{\natexlab{a}}.
\newblock \doarXiv{2401.15397}

\bibitem[{{Zhang} {et~al.}(2022){Zhang}, {Xu}, {Wang}, {Jing}, {Liu}, {Zhu}, \&
  {Jiang}}]{Zhang2021}
{Zhang}, C.-P., {Xu}, J.-L., {Wang}, J., {et~al.} 2022, Research in Astronomy
  and Astrophysics, 22, 025015, \dodoi{10.1088/1674-4527/ac3f2d}

\bibitem[{{Zhang} {et~al.}(2023){Zhang}, {Jiang}, {Zhu}, {Pan}, {Cheng}, {Liu},
  {Zhu}, {Sun}, \& {FAST Collaboration}}]{Zhang2023}
{Zhang}, C.-P., {Jiang}, P., {Zhu}, M., {et~al.} 2023, Research in Astronomy
  and Astrophysics, 23, 075016, \dodoi{10.1088/1674-4527/acd58e}

\bibitem[{{Zhang} {et~al.}(2024{\natexlab{b}}){Zhang}, {Zhu}, {Jiang}, {Cheng},
  {Wang}, {Wang}, {Xu}, {Liu}, {Yu}, {Qian}, {Yu}, {Ai}, {Jing}, {Xu}, {Liu},
  {Guan}, {Sun}, {Yang}, {Huang}, {Hao}, \& {FAST Collaboration}}]{Zhang2024}
{Zhang}, C.-P., {Zhu}, M., {Jiang}, P., {et~al.} 2024{\natexlab{b}}, SCIENCE
  CHINA Physics, Mechanics \& Astronomy, 67, 219511,
  \dodoi{https://doi.org/10.1007/s11433-023-2219-7}

\bibitem[{Zhang {et~al.}(2010)Zhang, Chen, \& Liang}]{Zhang2010}
Zhang, Z.-M., Chen, S., \& Liang, Y.-Z. 2010, Analyst, 135, 1138,
  \dodoi{10.1039/B922045C}

\bibitem[{{Zhou} {et~al.}(2023){Zhou}, {Zhu}, {Yang}, {Yu}, {Yuan}, {Jiang}, \&
  {Xi}}]{Zhou2023}
{Zhou}, R., {Zhu}, M., {Yang}, Y., {et~al.} 2023, \apj, 952, 130,
  \dodoi{10.3847/1538-4357/acdcf5}

\bibitem[{{Zhu} {et~al.}(2021){Zhu}, {Yu}, {Wang}, {Xu}, {Du}, {Yuan}, {Wang},
  {Jing}, {Ai}, \& {Jiang}}]{Zhu2021}
{Zhu}, M., {Yu}, H., {Wang}, J., {et~al.} 2021, \apjl, 922, L21,
  \dodoi{10.3847/2041-8213/ac350a}

\bibitem[{{Zuo} {et~al.}(2022){Zuo}, {Yang}, {Wang}, {Staveley-Smith}, {Lin},
  {For}, {Westmeier}, {Wang}, {Spekkens}, {Kilborn}, {Ivy Wong}, {Li},
  {Lee-Waddell}, {Catinella}, {Ho}, {Koribalski}, {Lee}, \& {Zhu}}]{Zuo2022}
{Zuo}, P., {Yang}, D., {Wang}, J., {et~al.} 2022, Research in Astronomy and
  Astrophysics, 22, 095016, \dodoi{10.1088/1674-4527/ac7f86}

\end{thebibliography}

\begin{appendix}




\renewcommand{\thesection}{Appendix}

\section{Velocity Comparison with GALFA-\HI} \label{sec:a}
\begin{figure}[H]
        \centering
	\includegraphics[width=1.\columnwidth]{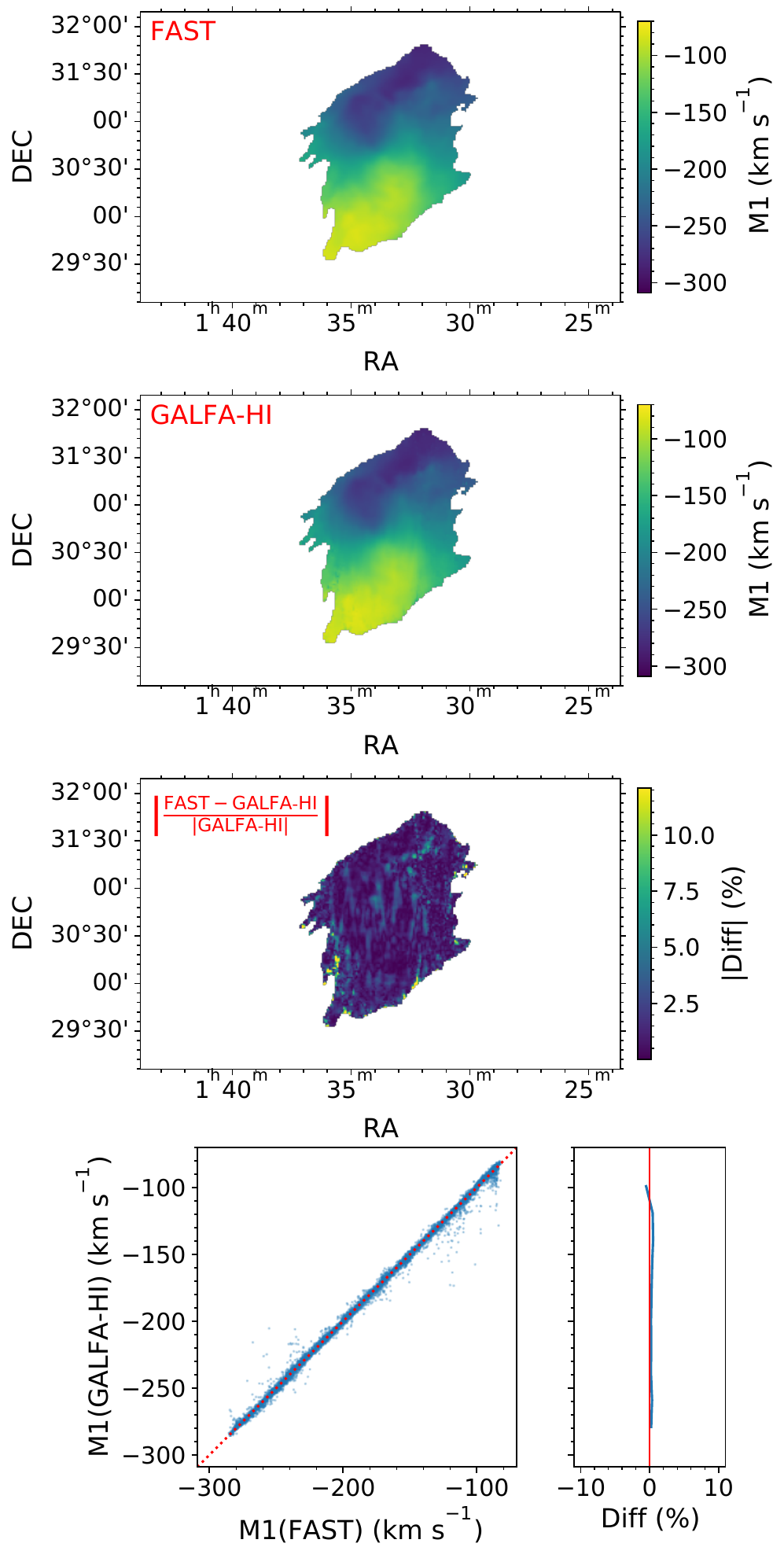}
	\caption{Similar to the right panels of Fig.~\ref{fig:m33-1}, but for comparison of data from FAST (this work) and Arecibo (GALFA-\HI) without velocity adjustment.}
	\label{fig:m33-GALFA}
\end{figure}

In Section~\ref{sec:ages}, we found that there is a systematic offset of about $-4.75$ km s$^{-1}$ between the FAST and AGES data. After accounting for this offset, the Moment-1 comparison displayed consistency, as shown in Fig.~\ref{fig:m33-1}. In this section, we further expand our comparison by comparing the data from FAST to the data from the Galactic Arecibo L-band Feed Array \HI\ Survey \citep[GALFA-\HI,][]{Peek2018}. GALFA-\HI\ also utilizes the Arecibo telescope but has a higher velocity resolution of approximately $0.7$ km s$^{-1}$. Fig.~\ref{fig:m33-GALFA} illustrates the Moment-1 comparison between FAST and GALFA-\HI. The median difference is approximately $0.52$ km s$^{-1}$, which is smaller than the difference between FAST and AGES. This suggests that the velocity resolution may contribute to this systematic veloctiy offset.

\end{appendix}

\end{multicols}
\end{document}